\documentclass[usenatbib]{mn2e}
\usepackage{graphicx}
\usepackage{amsmath,amssymb}
\usepackage{aas_macros}
\usepackage{rotating}
\usepackage{pdflscape}
\newcommand{\el}[2]{\ensuremath{^{#1}\mathrm{#2}}}
\newcommand{\Mo}{\rm{M}_\odot}
\newcommand{\Lo}{\rm{L}_\odot}

\newcommand{\Teff}{$T_{\rm{eff}}$}
\long\def\symbolfootnote[#1]#2{\begingroup%
\def\thefootnote{\fnsymbol{footnote}}\footnote[#1]{#2}\endgroup}
\usepackage{colortbl}%

\begin{document}

\title[Diagnostics of Stellar Modelling]{Diagnostics of Stellar Modelling from Spectroscopy and Photometry of Globular Clusters}

\author[Angelou et al.]
{\parbox{\textwidth}{George C. Angelou$^{1,2,3}$\thanks{E-mail: angelou@mps.mpg.de}, 
Valentina D'Orazi$^{2,4,5}$,
Thomas N. Constantino$^2$, \\
Ross P. Church$^6$,
Richard J. Stancliffe$^7$,
and John C. Lattanzio$^2$}
\vspace{0.4cm}\\
$^1$Max Planck Institut f\"ur Sonnensystemforschung, Justus-von-Liebig-Weg 3, 37077 G\"ottingen, Germany\\
$^2$Monash Centre for Astrophysics,  School of Physics and Astronomy, Monash University,  Clayton,  VIC 3800,  Australia.\\
$^3$Stellar Astrophysics Centre, Department of Physics and Astronomy, Aarhus University, Ny Munkegade 120, 8000 Aarhus C, Denmark\\
$^4$Department of Physics and Astronomy, Macquarie University, Sydney, NSW 2109, Australia.\\
$^5$INAF- Osservatorio Astronomico di Padova, Vicolo dell’Osservatorio 5, 35122, Padova, Italy \\
$^6$Department of Astronomy and Theoretical Physics, Lund Observatory, Box 43, SE-221 00 Lund, Sweden. \\
$^7$Argelander-Institut f\"{u}r Astronomie, Universit\"{a}t Bonn, Auf dem H\"{u}gel 71, D-53121 Bonn, Germany\\
}

\date{Accepted . Received ; in original form }

\pagerange{\pageref{firstpage}--\pageref{lastpage}} \pubyear{2014}

\maketitle

\label{firstpage}

\begin{abstract}

We conduct a series of comparisons between spectroscopic and photometric observations of globular clusters
and stellar models to examine their predictive power. Data from medium-to-high resolution spectroscopic surveys of lithium allow us to investigate first dredge-up
and extra mixing in two clusters well separated in metallicity. Abundances at first dredge-up are satisfactorily reproduced but there is preliminary evidence to suggest that the models overestimate the luminosity at which the surface composition first changes in the lowest-metallicity system. Our models also begin extra mixing at luminosities that are too high, demonstrating a significant discrepancy with observations at low metallicity. We model the abundance changes during extra mixing as a thermohaline process and 
determine that the usual diffusive form of this mechanism cannot simultaneously reproduce both the carbon and lithium observations. 
Hubble Space Telescope photometry provides turnoff and bump magnitudes in a large number of globular clusters and offers the opportunity to better test stellar modelling as function of metallicity.   We directly compare the predicted main-sequence turn-off and bump magnitudes as well as the distance-independent parameter $\Delta M_V ~^{\rm{MSTO}}_{\rm{bump}}$. We require 15 Gyr isochrones to match the main-sequence turn-off magnitude in some clusters and cannot match the bump in low-metallicity systems. 
Changes to the distance modulus, metallicity scale and bolometric corrections may impact on the direct comparisons  
but $\Delta M_V ~^{\rm{MSTO}}_{\rm{bump}}$, which is also underestimated from the models, can only be improved through changes to the input physics. 
Overshooting at the base of the convective envelope with an efficiency that is metallicity dependent is required to 
reproduce the empirically determined value of $\Delta M_V ~^{\rm{MSTO}}_{\rm{bump}}$.

\end{abstract}

\begin{keywords}
stars:abundances, evolution, interiors, Population II.  
\end{keywords}

%%%%%%%%%%%   Introduction  %%%%%%%%%%%%%%%%
\section{Introduction} \label{sec:intro}
Globular clusters (GCs) host mono-metallic\footnote{with a few exceptions, such as e.g., Omega Cenaturi \citep{2010ApJ...722.1373J}, M22 \citep{2009A&A...505.1099M}, M54 \citep{2010A&A...520A..95C}, M2 \citep{2014MNRAS.441.3396Y}}, almost coeval stellar populations that have been studied extensively, both photometrically and spectroscopically. In spite of their multiple populations (see \citealt{2012A&ARv..20...50G}), these systems offer well constrained tests for stellar evolution theory. Their colour magnitude diagrams (CMD, hence photometry) provide two clear evolutionary diagnostics for the stellar models -- the main-sequence turn-off (MSTO) magnitude and the magnitude of the bump in the red giant branch (RGB) luminosity function (LF bump or bump hereinafter). 
The MSTO magnitude serves as the classic indicator for cluster age, and by fitting this region of the CMD, modellers can 
take solace in the fact their codes provide a reasonable approximation to the stellar physics during the early phases of evolution.  
The bump magnitude, on the other hand, reveals information about the depth of first dredge-up (FDU) and stellar mixing -- it is a  probe of internal processes that are significant in later stages of evolution. Spectroscopy complements (and vice-versa) these photometric studies by providing further quantitative details of the internal mixing processes. It is our interest in stellar abundances
and mixing processes, particularly during the RGB ascent,  that motivate this study.

Mixing in RGB stars is characterised by two distinct events; the well-understood FDU and
an additional mixing episode that operates over a longer timescale during a more advanced phase of RGB evolution (``extra mixing" hereinafter).
Only mixing during FDU is predicted by standard stellar theory \citep{1967ApJ...147..624I}. 
In low-mass stars ($M \lesssim 1 \ \Mo$), deep convective motions develop once the star becomes a giant and these can penetrate into regions that have previously experienced partial hydrogen burning. 
Material enriched in \el{4}{He}, \el{14}{N} and \el{13}{C} is mixed through the
convection zone increasing the prevalence of these nuclei at the stellar surface. Conversely, \el{7}{Li} and \el{12}{C} abundances decrease as they are diluted into the extending envelope. In solar-metallicity stars the \el{12}{C}/\el{13}{C} ratio falls from the solar value of  $\approx 90$ to $\approx 30$ after the FDU, whilst \el{7}{Li} is depleted by a factor of $\approx 20$. Up to this point the theoretically predicted changes are in good agreement with observations.

It is clear from observations that after FDU the surface composition is further altered during RGB evolution \citep{1991ApJ...371..578G, 2000A&A...354..169G, 2003PASP..115.1211S, 2003ApJ...585L..45S, 2004MmSAI..75..347W,  2008AJ....136.2522M}. 
This second mixing event sees the cycling of hydrogen burning products into the convective envelope. From a post-FDU value of approximately 30, the \el{12}{C}/\el{13}{C} ratio reduces to 
$\approx 15$ in solar-metallicity stars and to the equilibrium value of $\approx 4$ in metal-poor stars. Lithium is essentially destroyed at all metallicities\footnote{Save for the small number of Li-rich giants discovered. See \citep{2015ApJ...801L..32D} for example.}. These changes are not predicted by standard stellar theory.

 The onset of extra mixing seemingly coincides with the bump in the luminosity function of GCs. The mixing is therefore associated with the advance of the hydrogen-burning shell into the composition discontinuity left behind by the deepest extent of FDU. The internal process(es) responsible for the surface changes are thought not to manifest themselves until after the bump because the composition discontinuity, and hence discontinuity in the mean molecular weight ($\mu$) gradient, acts as a barrier to any extra mixing process \citep{1979ApJ...229..624S, 1998A&A...332..204Cgr}.

In this study we are concerned with how well stellar models reproduce the key properties (magnitude/luminosity onset, abundance changes) of
several evolutionary features experienced during the late main-sequence and RGB phase of evolution. These include:
\begin{itemize}
\item the MSTO magnitude,
\item the LF bump magnitude,
\item the difference between the MSTO and bump magnitude,
\item the FDU magnitude,
\item the abundance changes associated with FDU,
\item the abundance changes associated with extra mixing. 
\end{itemize}

Observations of lithium in GC stars are used to constrain FDU and the extra mixing event.  Matching the magnitude of the LF bump has, in the past, been used as a measure of the adequacy of stellar models (e.g., \citealt{ 1985ApJ...299..674K, 1990A&A...238...95F, 1991A&A...244...95A,1997MNRAS.285..593C, 1997MNRAS.290..515C, 1999ApJ...518L..49Z, 2002PASP..114..375S,2003A&A...410..553R,2006ApJ...641.1102B,2006A&A...456.1085M, 2010A&A...510A.104M, 2010ApJ...712..527D, 2011A&A...527A..59C}). Here we draw upon some of the methodology of these studies and compare to recently available large data sets. 

We first turn our attention to spectroscopic surveys of lithium abundances in GCs and focus, in particular, on two clusters that are separated in metallicity by factor of ten.  These systems provide a detailed examination of FDU and extra mixing. 
In order to better sample the metallicity distribution, we utilise Hubble Space Telescope (HST) photometry to determine how well stellar models reproduce the MSTO magnitude, bump magnitude and the difference between the two (which is independent of distance and reddening) in a large number of systems. These three comparisons allow us to investigate the uncertainties introduced from the metallicity scale, distance modulus and bolometric corrections. We determine how changes to the input physics impact upon the level of agreement between observations and models. We conclude with a spectroscopic analysis of extra mixing and the ability of the thermohaline mechanism to simultaneously account for the depletion of carbon and lithium as a function of luminosity. Such a demand is a stringent test of any extra mixing process.  
\section{Overview of the Stellar Models} \label{modov}

The stellar models in this work are calculated with 
MONSTAR (the Monash version of
the Mt. Stromlo evolution code; see \citealt{2008A&A...490..769C}).
Opacities are provided at the high temperature end by the OPAL Rosseland mean opacity tables \citep{1996ApJ...464..943I}.
Our standard procedure is to employ tables that are variable in C and O content (OPAL type-2) based on the solar heavy element mixture specified by \citet[GN93]{1993oee..conf...15G}. The code also utilises additional (OPAL type-1) tables with a fixed metal distribution when required. 
For this work we have generated fixed metal distribution tables with the \citet[AGSS09]{2009ARA&A..47..481A} solar mixture, and alpha-element enhancements of  [$\alpha$/Fe]$= 0.2$ and [$\alpha$/Fe]$= 0.4$ of the AGSS09 abundances. 
Below $10^4$K  opacity tables from \citet{2009A&A...494..403L} with variable C and N content are used (see \citealt{simthesis} and \citealt{2008A&A...490..769C} for further details).

MONSTAR ordinarily only follows those species that are significant energetically. A seven species network (\el{1}{H}, \el{3}{He}, \el{4}{He}, \el{12}{C}, \el{14}{N}, \el{16}{O}, as well as a seventh pseudo-element that ensures baryon conservation) is sufficient to include  feedback on the structure from the nuclear energy generation. We have extended the network for this study so that the 
 evolution of \el{7}{Be}, \el{7}{Li} and \el{13}{C} are now followed; with the necessary changes to temporal and spatial resolution criteria to follow these fragile species (see \citealt{2015MNRAS.446.2673L} for the importance of these criteria). The key reactions and 
source of each adopted rate can be found in Table \ref{tab:rates}. Note that we employ the \el{14}{N}(p, $\gamma$)\el{15}{O} reaction rate provided by Champagne (private communication) which is consistent with that given by  \citet{2011RvMP...83..195A} and the LUNA collaboration. \citet{2011ApJ...729....3P} discuss the consequences of the new rate for low-mass stellar evolution, including extra mixing. 
\begin{table}
\centering

\begin{tabular}{ll} 

  \hline \hline
 Reaction Rate &  Source \\
  \hline
\el{1}{H}(p, e$^+ \nu_{\rm{e}})$\el{2}{H} &  \citet{1983ARAA..21..165H} \\
\el{3}{He}(\el{3}{He}, 2p)\el{4}{He} &  \citet{1988ADNDT..40..283C} \\
\el{3}{He}(\el{4}{He}, $\gamma$)\el{7}{Be} &  \citet{1988ADNDT..40..283C} \\
\el{12}{C}(p, $\gamma$)\el{13}{N} &  \citet{1988ADNDT..40..283C} \\
\el{14}{N}(p, $\gamma$)\el{15}{O} &  Champagne (2004, private comm) \\ 
\el{7}{Be}(e$^-$, $\nu_{\rm{e}}$)\el{7}{Li} &  Reaclib electron capture database \\
\el{7}{Be}(p, $\gamma$)\el{8}{B} &  \citet{1999NuPhA.656....3A} \\
\el{7}{Li}(p, \el{4}{He})\el{4}{He} &  \citet{2004ADNDT..88..203D} \\
\el{13}{C}(p, $\gamma$)\el{14}{N} &  \citet{1999NuPhA.656....3A}\\
\hline
\end{tabular}
\caption{Key reaction rates used in MONSTAR.}
\label{tab:rates}
\end{table}

Convective energy transport in MONSTAR is treated according to the mixing length theory (MLT, \citealt{1958ZA.....46..108B}). Mixing of the chemical species is calculated using a diffusion equation \citep{2008A&A...490..769C}. 
With an assumed helium content of Y$=0.2485$ and heavy element mixture specified by GN93, the MLT parameter $\alpha_{MLT} =1.75$  best reproduces the solar model at the Sun's current age. 
A value of $\alpha_{MLT} =1.69$ is required when the AGSS09 abundances are adopted. 
In the models presented here, we employ the Schwarzschild criterion and where specified, allow for non-locality by including a prescription for diffusive overshoot. We follow the procedure of \citet{1997A&A...324L..81H} who parameterised the numerical simulations of convection by \citet{1996A&A...313..497F}. Those results indicated an exponential decay in velocity of the overshooting material. In analogy to the pressure scale height, $H_P$, a `velocity scale height', $H_v$, is defined such that  

\begin{equation} \label{eqn:os1}
H_v = f_{os} H_P
\end{equation} 
where $f_{os}$ is a scaling factor which we vary between $f_{os}=0.0-0.1$ in this study. The resulting equation for the diffusion coefficient is then
\begin{equation}
D_{os} = D_0 \ e^{\frac{-2z}{H_v}} \label{eqn:os3}
\end{equation}
where $D_0$ is the diffusion coefficient at the last convective point and \textit{z} is distance from the convective boundary.

Our implementation of the thermohaline mechanism uses the formulation
developed by \citet{1972ApJ...172..165U} and 
\cite*{1980A&A....91..175K}, in which
thermohaline
mixing is modelled as a diffusive process. 
This prescription has been employed in previous work by  
\citet{2007A&A...467L..15C,2007A&A...476L..29C}, \citet{2009MNRAS.396.2313S}, \citet{2010MNRAS.403..505S},
\citet{2010A&A...522A..10C}, and
\citet{2011ApJ...728...79A, 2012ApJ...749..128A}

The equation for the diffusion coefficient is:
\begin{equation} 
D_t =  C_t \,  K  \left({\varphi \over \delta}\right){- \nabla_\mu \over
(\nabla_{\rm ad} - \nabla)} \quad \hbox{for} \;  \nabla_\mu < 0 ,
\label{eqn:dt}
\end{equation}
where $\varphi = (\partial \ln \rho / \partial \ln \mu)_{P,T}$, 
$\delta=-(\partial \ln \rho / \partial \ln T)_{P,\mu}$, 
$\nabla_{\rm{\mu}} = (\partial \ln \mu / \partial \ln P)$, 
$\nabla_{\rm{ad}}=(\partial \ln T / \partial \ln P)_{\rm{ad}}$, 
$\nabla = (\partial \ln T / \partial \ln P)$, 
\textit{K} is the thermal diffusivity and $C_t$ is a dimensionless free
parameter.
In this diffusive theory, $C_t$ is related to the aspect ratio, $\alpha$, of the 
thermohaline fingers (assumed to be cylindrical) by

\begin{equation} 
C_t = {8 \over 3} \pi^2 \alpha^2.
\label{eqn:ct}
\end{equation}

The mechanism is elegant in that the depth of mixing is set by the stellar structure
resulting in only one free parameter. An empirically derived value of $C_t = 1000$ can reproduce abundance patterns in both globular cluster
stars \citep{2011ApJ...728...79A,2012ApJ...749..128A} and field stars \citep{2010A&A...522A..10C} as well as
 the dichotomy between carbon-normal and carbon-enhanced metal-poor stars \citep{2009MNRAS.396.2313S}.   
\citet{2010ApJ...723..563D} and \citet{2010A&A...521A...9C} prefer a lower value of $C_t = 12$ on theoretical grounds, which is also supported by 3D hydrodynamical models that suggest the mixing is inefficient on the RGB  \citep{2011ApJ...727L...8D, 2011ApJ...728L..29T, 2013ApJ...768...34B}. Parametrising extra mixing in this form and comparing to observations can still tell us much about the transport of material in the stars. Similar conclusions could be reached had we elected to include a phenomenological mixing model \citep{2003ApJ...593..509D}. 
Note that in such models the depth of mixing is usually specified by a constant shift in either
mass or temperature from the hydrogen-burning shell. An additional free parameter sets the mixing speed. This is a less physically motivated but similar configuration to that used here. 

In order to compare to observations, we convert our stellar models from luminosity space to absolute visual magnitude ($M_V$) throughout.
This requires a $V$ band bolometric correction for each stellar model calculated. 
Theoretical model atmospheres from \citet[][ATLAS9]{1997A&A...318..841C} provide the necessary tables of bolometric corrections and we determine the most appropriate value by using a cubic spline to interpolate in composition, surface gravity and \Teff .

\section{First Analysis of the Spectroscopic Data}

\subsection{Observations of Lithium in Globular Clusters}
The lithium abundances used to constrain our models come from medium-to-high resolution surveys of GCs. 
Although far more complex than their classic simple stellar population archetype, GCs are still useful testbeds of stellar theory due to their well populated colour-magnitude diagrams (CMDs). These systems host multiple stellar populations spanning relatively small differences in age and are characterised by internal variations in their light-element (e.g., C, N, Na, O, Al, see \citealt{2012A&ARv..20...50G} and references therein for a recent review on multiple populations) and \el{4}{He} content \citep{2010A&A...517A..81G,2011A&A...534A.123G,2012A&A...539A..19G, 2014ApJ...785...21M}.

In addition to these primordial abundance variations, the constituent stars undergo \textit{in situ} composition changes such as those experienced on the RGB. \textit{In situ} mixing manifests itself as a function of luminosity and is easily discerned.
The carbon and nitrogen abundances, in particular, have been extensively measured.  The \el{12}{C}/\el{13}{C} ratio is also a useful tracer of mixing. It scrutinises FDU robustly \citep{1975MNRAS.170P...7D,1976ApJ...210..694T, 1994A&A...282..811C}, but saturates rapidly once extra mixing begins.  
Measuring \el{12}{C}/\el{13}{C} requires spectra with medium-to-high resolution and signal-to-noise ratios.

Lithium is a very useful probe of mixing and a sensitive gauge of temperature because it is destroyed at 2 MK.  It can therefore  provide an indication of the mixing efficiency of both FDU and the extra-mixing event. It also has the advantage that its abundance determination is reliable: it is derived from the the Li~{\sc i} resonance doublet at 6707.78\AA , rather than from molecular bands which is the case for C and N.  

We have compiled data from four medium-to-high resolution studies that focus on lithium abundances in globular cluster giants. 
Data from all surveys have typical uncertainties of roughly $\pm$0.1
in A(Li) and $\pm$0.03 mag.
The aforementioned internal (i.e., star-to-star) errors in Li abundances
are due to a combination of uncertainties in
  equivalent width measurements, continuum placement, signal-to-noise
ratios of the spectra and atmospheric parameters
(effective temperatures, micorturbulence, gravity and metallicity, with the first
contribution being the largest).
We refer to \citet{2014ApJ...791...39D} for an extensive discussion of this topic.
Our selected sample of clusters spans a large range in metallicity and we expect to identify any such trends present. 
In Table \ref{tab:clusters} we list the clusters with Li abundances determined and their general properties. The metallicities are as provided in \citet{2013ApJ...766...77N} and the \citet{1996AJ....112.1487H} catalogue (2010 edition).

\begin{table*}
\begin{center}
\begin{tabular}{ccccccccc} 

  \hline \hline
 ID &  Messier ID & Relative & Luminosity & [Fe/H] & RGB$_{\rm{bump}}$   &  FDU    &  Number of & Source \\
    &             & Age      & ($M_V$)    &        &  ($M_V$)            & ($M_V$) &   Targets &  \\
  \hline

NGC 6121 & M4  & $0.91$ & $-7.19$ & $-1.10$ & $0.45$ & $3.65$   & $87$ & \citet{2011MNRAS.412...81M} \\
NGC 2808 & --  & $0.74$ & $-9.39$ & $-1.18$ & $0.65$  & -- &  $68$ & \citet{2015arXiv150305925D} \\
NGC 5904 & M5  & $0.81$ & $-8.81$ & $-1.29$ & $0.50$ & --  &$99$ & \citet{2014ApJ...791...39D} \\
NGC 362  & --  & $0.74$ & $-8.43$ & $-1.30$ & $0.57$ & --  & $67$ & \citet{2015arXiv150305925D} \\
NGC 6218 & M12 & $0.92$ & $-7.31$ & $-1.37$ & $0.78$ & --  & $63$ & \citet{2014ApJ...791...39D} \\
NGC 1904 & M79  & $0.87$ & $-7.86$ & $-1.58$ & $0.29$& --  &  $47 $ & \citet{2015arXiv150305925D} \\
NGC 6397 & --  & $0.99$ & $-6.64$ & $-2.10$ & $0.16$ & $3.3$   &$454$ &   \citet{2009AA...503..545L} \\
\hline
\end{tabular}
\caption{General properties of the clusters from which spectroscopic data is utilised in this work. Relative ages are as defined in \citet{2005AJ....130..116D}. }
\label{tab:clusters}
\end{center}
\end{table*}

\begin{figure}
\centering
 \includegraphics{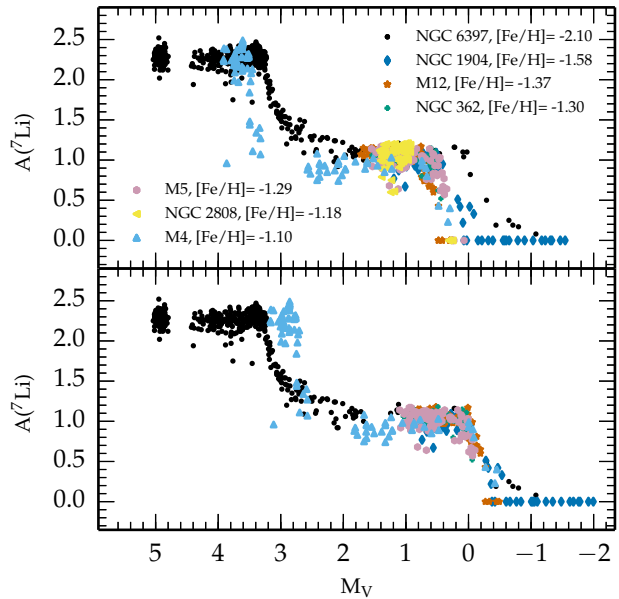}
\caption{Top Panel: A(\el{7}{Li}) as a function of absolute magnitude ($M_V$) for the clusters specified in the legend. Bottom Panel: As above but with an offset applied to the magnitude so that the LF bumps coincide. In this panel we omit data from NGC 2808 as none of the stars were observed in the luminosity range where Li depletion via extra mixing is expected to occur. }
 \label{fig:licl}
\end{figure}

The most metal-rich cluster in our sample, M4 ([Fe/H] $=-1.10$), and the most metal-poor, NGC 6397 ([Fe/H] $=-2.10$),  have subgiant branch lithium abundances consistent with the Spite plateau \citep[see also Figure 1]{1982A&A...115..357S}; a point discussed by the authors of the original surveys.  
In these two clusters the first instance of lithium depletion marks the onset of FDU.
The difference in metallicity causes stars in M4 to begin FDU approximately $\Delta M_V \approx 0.3$ magnitudes fainter than those in NGC 6397. This is because in metal-rich stars, the additional low-ionisation metals (i.e., Ca, Na, K, and Al) contribute to the higher opacity. All other things being equal, the metal-rich stars:
\begin{enumerate}
\item begin the inward migration of the convective envelope at a lower luminosity; and
\item develop deeper convective envelopes. 
\end{enumerate} 
compared with metal-poor stars.
As one might expect from deeper dredge-up, stars in M4 exhibit a post-FDU Li abundance that is generally lower than those in NGC 6397. 
However it is not certain that we are detecting a metallicity effect because systematic uncertainties may exist between the two studies.
The upper range of the Li abundances in M4 is consistent with the stars in NGC 6397. Furthermore, the significant change in abundances during FDU is due to the transition from shallow surface convection to a deep convective envelope. A slightly deeper convective envelope may not necessarily translate to a detectable difference in the surface abundance. The two studies found the same pre-FDU abundances which might suggest that the systematic errors are not significant and the differences in the mean pre-bump Li abundances are indeed due to metallicity. Systematic uncertainties that exist \textit{between} surveys, however, can only truly be minimised through a homogeneous study of both clusters (i.e., identical instrument, line lists, codes, methodology). Interestingly, those clusters self-consistently analysed by \citet{2014ApJ...791...39D} and \citet{2015arXiv150305925D} span a smaller metallicity range (see Table \ref{tab:clusters}) but share the same mean pre-bump Li abundance once observational uncertainties are taken into account.

The two clusters show noticeably different gradients of lithium depletion during FDU. Stars in M4 complete FDU over a much smaller luminosity range than those in NGC 6397. The abundance trend reflects the rate of advance of FDU which is different because the greater opacity allows the convective envelope to penetrate faster in the metal-rich regime. 

Figure \ref{fig:licl} also provides insight into the extra mixing process. The GCs in our sample have been surveyed across a large luminosity range that includes the LF bump. Because the depth of FDU is metallicity dependent, so too is this secondary mixing event. Metal-rich stars begin extra mixing at fainter magnitudes because the hydrogen shell is not required to advance as far before it encounters the homogenised region and removes the $\mu$ inversion that inhibits the mixing process. We note that such a metallicity trend is not evident from Figure \ref{fig:licl}. This clear in Table \ref{tab:clusters} which shows the statistically determined LF bump magnitude \citep{2013ApJ...766...77N} for each cluster sorted by metallicity. The metal-rich clusters do not necessarily begin extra mixing at fainter magnitudes as theory predicts. 
Uncertainty in the metallicity \footnote{Whilst there may be some uncertainty in the metallicity determination of these systems, it is a robust result that NGC 6397 and M4 differ by approximately a factor of 10 in their metal content.} and distance modulus determinations
and the role of multiple populations each contribute to the observed behaviour. Such disagreement with direct comparison of the evolutionary events is a common theme throughout this paper.

In the bottom panel of Figure \ref{fig:licl} we have applied a magnitude offset to each cluster so that the magnitude at which extra mixing begins is common to all clusters. In all clusters, irrespective of metallicity, extra mixing depletes lithium at a similar rate. Agreement is further improved when one considers that the abundances of the brightest three stars in NGC 6397 are upper limits. 
Such strong agreement is perhaps not unexpected given the fragility of lithium, but it also highlights a clear property (and constraint) of the physics that drives the mixing during this epoch.

\subsection{Surface Composition Changes During the Red Giant Branch Mixing Events} \label{gastuff}

\subsubsection{Shedding New Light on Previous Work with [C/Fe] and [N/Fe]}

\begin{figure} 
 \includegraphics{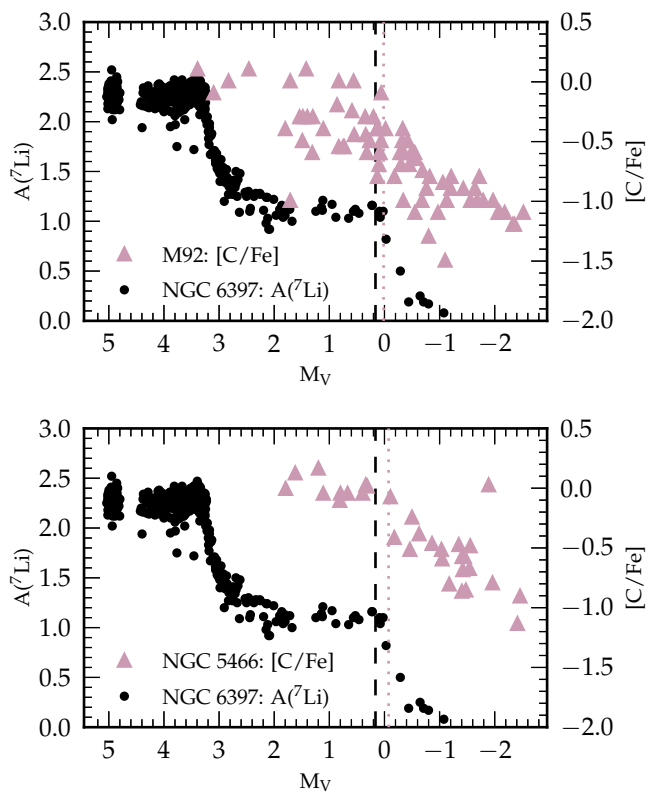}
\caption{Left Axis (both panels): Black circles denote A(\el{7}{Li}) as a function of magnitude for stars in NGC 6397 ([Fe/H]$=-2.1$). The dashed vertical line corresponds to the photometrically determined LF bump of the cluster (M$_{\rm{V}}=0.163$). 
Right Axis (top panel): Lavender triangles denote [C/Fe] as function of magnitude for stars in M92 ([Fe/H]$=-2.2$). The dotted vertical line corresponds to the photometrically determined LF bump of the cluster (M$_{\rm{V}}=0.016$).
Right Axis (bottom panel): Lavender triangles denote [C/Fe] abundance as function of magnitude for stars in NGC 5466 ([Fe/H]$=-2.2$). The dotted vertical line corresponds to the photometrically determined LF bump of the cluster (M$_{\rm{V}}=-0.075$). The LF bump for each cluster was determined by \citet{2013ApJ...766...77N}. }
 \label{fig:licfe}
\end{figure}

When trying to match the extra mixing event in M92 ([Fe/H] $=-2.2$), \citet{2012ApJ...749..128A} found that their models  underestimated the magnitude (hence overestimated the luminosity) of the LF bump by $\Delta M_V \approx 0.7$ mag. They were required to artificially deepen FDU significantly \footnote{The depth of FDU was extended from m$=0.368$ $\Mo$ to $m=0.320$ $\Mo$, where m is the enclosed mass.}  in their calculations to match the photometrically determined magnitude of the LF bump. An extension of the convective envelope was also 
required to match the LF bump in M15 (also [Fe/H] $=-2.2$).  In these clusters, it is unclear if the onset of extra mixing coincides with the photometric bump. One interpretation of the data is that surface abundance changes begin before the LF bump.  
Possible reasons for the discrepancy were given as uncertainties in the spectroscopy (e.g., combining data sets, determining abundances from molecular bands), difficulties in determining the luminosity of the bump at low metallicity, or that  extra mixing had initiated before the LF bump (which would prove to be a serious issue for stellar evolution). The authors noted that homogeneous lithium data would reveal the true behaviour of the cluster.

The three low-metallicity clusters studied by Angelou et al. (2012) [M92, M15 and NGC 5466, [Fe/H] $=-2.2$]  are yet to have their lithium abundances systematically measured. They do, however,  have comparable metallicity to the surveyed NGC 6397 ([Fe/H] $=-2.1$). In the upper panel of Figure \ref{fig:licfe} we plot [C/Fe] from M92 (right axis, lavender triangles) and A(Li) from NGC 6397 (left axis, black circles) as functions of absolute visual magnitude.   
Data for M92 are taken from \citet{2003PASP..115.1211S} who applied offsets to the studies by \citet{1982ApJS...49..207C}, \citet{1986PASP...98..473L} and
\citet{2001PASP..113..326B} in order to remove systematic
differences in abundance scales. Combining results from different studies, together with inferring abundances from molecular bands, may lead to uncertainties of up to 0.3 dex in the [C/Fe] data. Data for NGC 6397 are from \citet{2009AA...503..545L} and include their reanalysis of the survey by \citet{2009AA...505L..13G}. The dotted lavender line indicates the magnitude of the LF bump in M92 ($M_V=0.016$; \citealt{2013ApJ...766...77N}) and the black dashed line the magnitude of the LF bump in NGC 6397 ($M_V=0.163$; \citealt{2013ApJ...766...77N}). The magnitude at which extra mixing begins in the massive GC M92 is unclear from [C/Fe] from spectroscopy. The lithium decrease in NGC 6397, on the other hand, has a well defined starting magnitude that corresponds to its 
photometrically determined bump and the bump of the similarly metal-poor M92. 
In the lower panel we compare data from the less massive cluster NGC 5466. As per the panel above, [C/Fe] is denoted by lavender triangles with the scale provided on the right axis. \citet{2013ApJ...766...77N} determined the  magnitude of the LF bump in this cluster to be $M_V=-0.075$.  In NGC 5466 and NGC 6397 the respective magnitudes of the LF bump and onset of surface abundance changes (due to extra mixing) agree quite well. 

Unlike M92, for which the data are a combination of multiple surveys, data for NGC 5466 are homogeneous (they are taken from a single study by \citealt{2010AJ....140.1119S}). It may be that the combination of multiple data sets clouds the true cluster behaviour. If the uncertainty surrounding M92 is purely due to observational spread, then results from homogeneous surveys such as APOGEE will shed light on this issue.     
However, we note that both NGC 6397 and NGC 5466 are much less massive than M92 and \citet{2010A&A...516A..55C} and \citet{2014ApJ...791...39D} have demonstrated how clusters with greater mass can exhibit a greater spread in their primordial abundances.

Because the Li abundances suggest that the beginning of extra mixing in low-metallicity clusters does indeed coincide with the magnitude of the photometrically derived LF bump, then it is clear that the models presented in \citet{2012ApJ...749..128A} were underestimating the magnitude of the bump and of the onset of extra mixing.  This is irrespective of the chosen extra mixing mechanism and dependent on the physics of the stellar codes. 
Such an inconsistency was identified by  \citet{1985ApJ...299..674K} and \citet{1990A&A...238...95F}. We explore the extent of this discrepancy in \S4.

\subsubsection{Comparison with Models: Elucidating the Red Giant Branch Mixing Events at Globular Cluster Metallicities}

\begin{figure*}
 \includegraphics{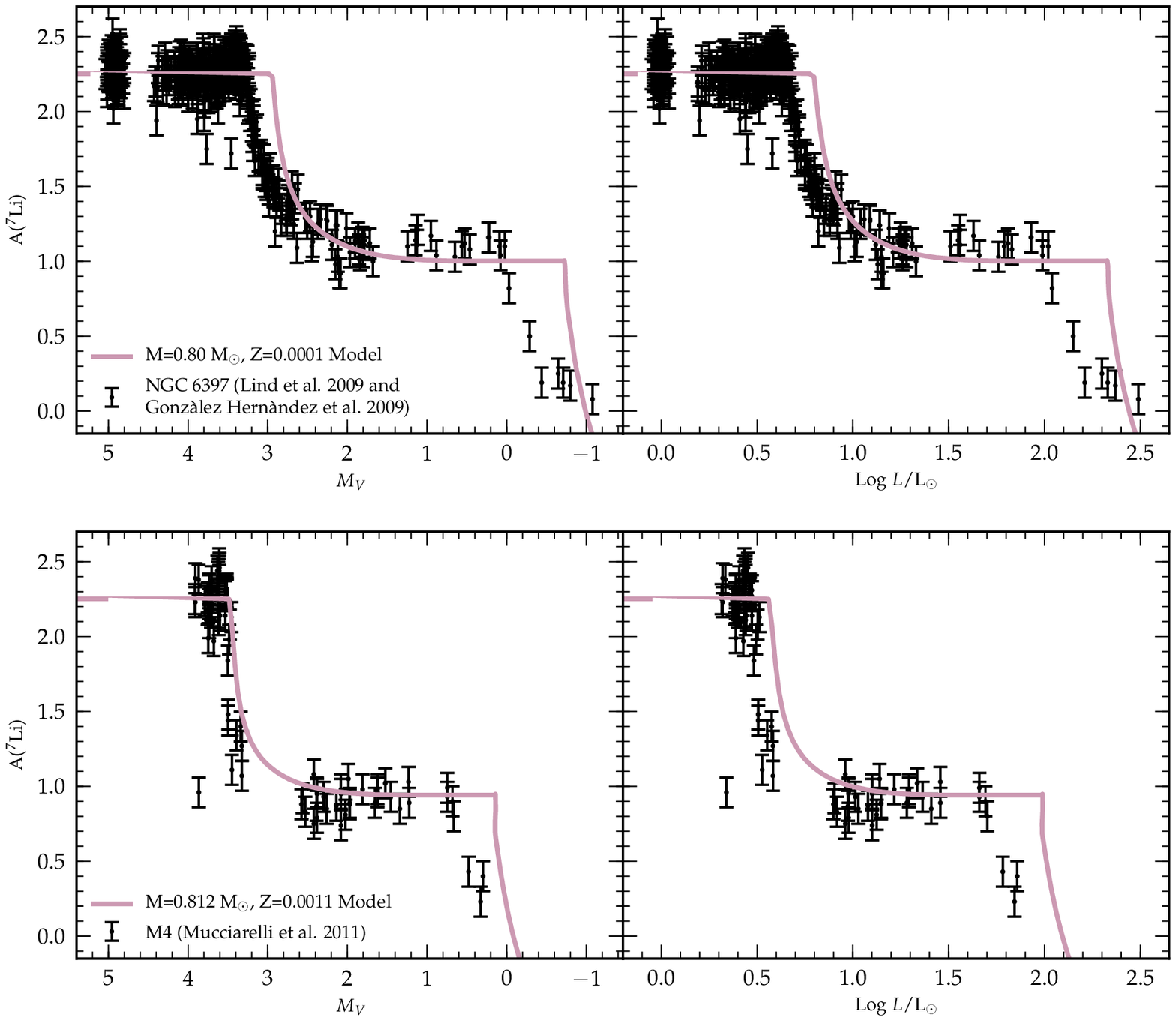}
\caption{Top panels: A(Li) as a function of absolute magnitude (left) and independently determined luminosity (right) in the globular cluster NGC 6397. Data are taken from \citet{2009AA...503..545L} and \citet{2009AA...505L..13G}. Bottom panels:  A(Li) as a function of absolute magnitude (left) and independently determined luminosity (right) in the globular cluster M4. Data for this cluster are taken from \citet{2011MNRAS.412...81M}. In each panel the solid lavender curve denotes a first approximation model for the respective cluster (model parameters are listed in Table \ref{tab:models} ).  }
 \label{fig:clustermods}
\end{figure*}

\begin{table}

%\begin{center}
\centering
\begin{tabular}{ccc} 

  \hline \hline
&   NGC 6397 & M4 \\
\hline
Mass ($M_{\odot}$) & 0.80 & 0.812 \\
Z & $0.00011$ & $0.0011$ \\
X(\el{4}{He}) & $0.24989$ & $0.2489$ \\
X(\el{1}{H}) &$0.75$ & $0.75$  \\
X(\el{12}{C}) &$1.99 \times 10^{-5}$ &  $1.99 \times 10^{-4}$ \\
X(\el{14}{N}) &  $5.83 \times 10^{-6}$ & $5.83 \times 10^{-5}$ \\ 
X(\el{16}{O}) & $4.82 \times 10^{-5}$ & $4.82 \times 10^{-4}$ \\
X(\el{7}{Li}) & $9.39 \times 10^{-10}$ & $9.39 \times 10^{-10}$ \\
$\alpha_{\rm{MLT}}$ & $1.75$ & 1.75 \\
$C_t$ & $1000$ & $1000$ \\
TO Age (Gyr) & 12.0 & 12.0 \\
Opacity Mixture &  AGSS09 &  AGSS09 \\
\hline
\end{tabular}
\caption{Model details used to fit the clusters NGC 6397 and M4 in Figure \ref{fig:clustermods}. Note that we require the same initial X(\el{7}{Li}) despite the factor of ten difference in metallicity.}
\label{tab:models}
%\end{center}
\end{table}

Our sample of GCs includes the metal-poor cluster, NGC 6397 ([Fe/H] $\approx -2.10$), and the metal-rich, NGC 6121 (M4, [Fe/H] $\approx -1.10$), which  have been surveyed across a luminosity range that covers FDU through to the extra mixing event. Li abundance determinations for NGC 6397 are presented in the top row of Figure \ref{fig:clustermods} and for M4 in the bottom row. 
The data are plotted both as a function of  
absolute visual magnitude (left panels) and as a function of independently determined luminosity (right panels). The two brightness systems are employed as a check on possible systematic errors in our method (see below). We calculate stellar models for each cluster
(lavender curves) with the parameters specified in Table \ref{tab:models}  as a first approximation. 

The models for NGC 6397 underestimate the magnitude of both mixing events. The difference in magnitude at FDU ($\approx 0.3$ mag) is not as pronounced as for the LF bump ($\approx 0.7$ mag). In the models for the metal-rich cluster, M4, the onset of FDU is consistent with the observations whilst a discrepancy is present at the LF bump ($\approx 0.4$ mag). It is well documented that theoretical models underestimate the magnitude of the LF bump \citep{1990A&A...238...95F,2006ApJ...641.1102B, 2010ApJ...712..527D,2011A&A...527A..59C}, however this is the first time a disagreement at FDU has been identified. The recent availability of lithium observations, which probe both RGB mixing events, allows for an investigation of FDU across a range of metallicities. Previous investigations of mixing during the RGB have relied on [C/Fe] and [N/Fe] which are of limited use for the FDU event because these species change very little during FDU at metallicities typical of GCs. Furthermore, their abundance determination (usually from molecular bands) is less robust than for Li. We have seen that in the lowest metallicity clusters there is an intrinsically large spread in C and N which makes identifying the onset of mixing by spectroscopy difficult. \citet{1994A&A...282..811C, 1995ApJ...453L..41C} has used both  \el{12}{C}/\el{13}{C} and Li to test extra mixing due to rotational instabilities. Model comparisons to M4 and Halo stars with NGC 6397 metallicity were presented with their predicted FDU magnitudes comparable or slightly brighter than ours. Models by \citet{2014ApJ...797...21P} determine that reproducing the onset of extra mixing is also a problem for carbon-enhanced metal-poor stars. 

\subsection{Bolometric Corrections}
It is concerning how much the models overestimate the brightness of the mixing events, especially the LF bump at low metallicity. 
It is prudent to first ascertain whether the magnitude offset between the theoretical and observationally determined LF bump  
is simply a result of a systematic error introduced through the conversion from luminosity to absolute magnitude.
The fact that the magnitude difference at FDU is not as pronounced as for the LF bump does not rule out a conversion problem. The bolometric corrections are functions of metallicity, \Teff\ and surface gravity and thus vary throughout evolution.
\citet{2002PASP..114..375S} have found that by changing the model atmosphere sets, bolometric corrections can differ by up to 0.1 magnitudes. This is not enough to account for the $\approx 0.3$ M$_{\rm{V}}$ offset between the models and observations at FDU.
 Even if we were to apply bolometric corrections from an empirically calibrated set of model atmospheres (e.g., \citealt{2000AJ....119.1448H}), it does not rule out the presence of systematic errors in our conversion. To reduce this source of error we use independently converted visual magnitudes calculated by \citet{2008A&A...490..777L} in their study of NGC 6397. 
Their method, similar to ours described above, converts visual magnitude to luminosity by applying a calibration from \citet{1999A&AS..140..261A}. This calibration, too, is a function of (independently determined) metallicity and \Teff. Magnitude was converted to luminosty through a 13.5 Gyr isochrone for the cluster \citep[which placed their stars in the mass range $0.78-0.79 \ \Mo$; similar to the $0.8 \ \Mo$ that we have modelled here]{2005ApJ...619..538R}.  The \citet{2008A&A...490..777L} results are presented in the top right panel of Figure \ref{fig:clustermods}. Our stellar model is again denoted by the solid lavender curve and in this case, the luminosity is calculated directly from the equations of stellar structure. The left (abundances as a function of $M_V$) and right (abundance as a function of luminosity) panels look remarkably similar with the expected factor of 2.5 difference in the respective brightness scales. Mucciarelli (private communication) has also provided us with independently determined luminosities for M4 (Figure \ref{fig:clustermods}). The two brightness scales ($M_V$ and luminosity) yield the same behaviour in each cluster, thus systematic uncertainties in our conversion are unlikely responsible for the magnitude discrepancy between theory and observation.  

We also have no reason to believe that there is a systematic issue with bolometric corrections determined from model atmospheres at low metallicity. In fact one would expect the opposite to be true; bolometric corrections at high metallicity should be more uncertain. 
Synthetic colors perform quite well in the metal-poor regime, without significant deviations between metallicities of [Fe/H] $=-1.5$ and [Fe/H] $= -2.5$ (Casagrande, private communication). As the metallicity increases, however,  the choice of atomic line-lists becomes increasingly important. 
We note that microturbulence in low-metallicity model atmospheres is one possible source of systematic differences.  
Microturbulence can affect UV/blue wavelengths so if the velocities were to change in the low-metallicity regime, then the derived corrections would be systemically offset. As no evidence for such behaviour exists, a closer look at the physics of stellar modelling is required.  

\section{Analysis of the Photometric Data} 
The results from NGC 6397 and M4 raise a series of interesting questions that only a larger sample of GCs will help answer: 
\begin{enumerate}
\item Do stellar models reproduce the structure of high-metallicity stars better than low-metallicity stars? 
\item Do uncertainties in the distance modulus make comparison between theoretical and observed FDU and bump magnitudes (direct comparison hereinafter) too inconsistent to be meaningful? 
\end{enumerate}
Unlike FDU which can only be identified by spectroscopic determination of abundances, the LF bump is readily identified through the CMD and hence has been observed in many more clusters. By switching to photometric data a second evolutionary indicator in the main-sequence turn-off (MSTO) magnitude can also be employed as a check on the models.

Comparisons between models and photometric GC data are common in the literature. There are three well tried methods by which we can compare theoretical predictions with empirical measurements of the LF bump:
\begin{enumerate}
\item By comparing the parameter $\Delta V^{\rm{bump}}_{\rm{HB}} = V_{\rm{bump}}-V_{\rm{HB}}$; the V magnitude difference between the RGB bump and the horizontal branch at the RR Lyrae instability strip magnitude \citep{1990A&A...238...95F,1999ApJ...518L..49Z, 2002PASP..114..375S,2003A&A...410..553R,2006ApJ...641.1102B,2006A&A...456.1085M, 2010ApJ...712..527D}.
\\
\item By comparing the parameter $\Delta V^{\rm{MSTO}}_{\rm{bump}} = V_{\rm{MSTO}}-V_{\rm{bump}}$; the V magnitude difference between the RGB bump and the MSTO \citep{2011A&A...527A..59C}.  
\\
\item By comparing the photometrically determined absolute magnitude of the LF bump to that predicted by the models \citep{1985ApJ...299..674K, 1991A&A...244...95A, 2006ApJ...641.1102B, 2012ApJ...749..128A}. Each stellar model requires a bolometric correction to convert luminosity to $M_V$  whilst the distance modulus is required to convert the observed visual magnitude to absolute magnitude.
\end{enumerate}
The first two methods have the advantage of being independent of distance and reddening.  %and not affected by the zero point offset in the photometry.  
As discussed by \citet{2011A&A...527A..59C}, the first method does introduce some uncertainty with respect to the placement of the
observed HB level for GCs with blue HB morphologies and in theoretical predictions of the HB luminosity (dependent on each code and their predicted \el{4}{He} core mass at the \el{4}{He} ignition at the RGB tip). The third method introduces an extra source of uncertainty because it relies on accurate determination of the distance modulus (as well as suffering from uncertainty due to interstellar reddening).
In order to understand the level of error introduced from uncertainties in the distance modulus
we  employ both the second and the third method in our investigation of the RGB stellar models.

The natural method for comparing the key theoretical and photometric indicators of stellar evolution has 
been through fitting of isochrones to the CMD. In this study we calculate individual stellar models at distinct metallicities
and fit third order polynomials to create what are essentially coarse isochrones.  
Table \ref{tab:natafmodels} outlines our sampling of the GC metallicity range and indicates the corresponding initial masses 
for each metallicity that yield a MSTO age of 12 Gyr.

We note that it is common to require isochrones with ages greater than the age of the Universe to match the luminosity of the LF bump (see \citealt{2003A&A...410..553R} and \citealt{2011A&A...527A..59C}). As \citet{2011A&A...527A..59C} comprehensively discuss, the reason that such old ages are needed is because the underlying stellar models fail to reproduce the RGB bump brightness for an age appropriate to the GC as measured from its MSTO brightness. Thus there exists clear motivation to focus on the underlying stellar models and the included physics.

GCs are observed to have enhanced $\alpha$-element abundances compared to scaled-solar values and
their metallicities are often expressed in terms of the \textit{total metallicity}, [M/H]. This is defined by \citet{1993ApJ...414..580S} as
\begin{equation}
\rm{[M/H]=[Fe/H]}+\log (0.638 \times 10^{[\alpha /\rm{Fe]}} +0.362).
\end{equation}
This metallicity definition has been used in similar studies and is adopted here.

As we have mentioned, this type of comparison has been carried out extensively in the literature. 
In their analysis of the $\Delta V^{\rm{bump}}_{\rm{HB}}$ parameter, \citet{2003A&A...410..553R} find good agreement between theory and observations at higher metallicities but note significant discrepancies at low metallicity. Investigations of low-metallicity GCs include those by \citet{1990A&A...238...95F} and \citet{2010ApJ...712..527D}, who also used the parameter  $\Delta V^{\rm{bump}}_{\rm{HB}}$, to determine that at  [M/H] $\lesssim -1.7$, models underestimate the magnitude of the LF bump by at least 0.4 mag. \citet{2010ApJ...712..527D} also conducted tests on the effects of the microphysics and determined that models with $\alpha$-element and CNO-enhancements could not account for the discrepancy nor could the revised solar heavy-element mixture. \citet{2011A&A...527A..59C}, who employed the $\Delta V^{\rm{MSTO}}_{\rm{bump}}$ parameter, found that the theoretical bump was too bright by 0.2 magnitudes on average, but discrepancies of $\Delta M_V \gtrsim 0.4$ magnitudes in the lowest-metallicity clusters were within the uncertainties. 
In their test of the stellar microphysics, \citet{2006ApJ...641.1102B} have determined that the uncertainty in theoretical values for the LF bump magnitude varies with metallicity between $+0.13$ and $-0.12$ magnitudes at [Fe/H] $=-2.4$ and between
$+0.23$ and $-0.21$ magnitudes at [Fe/H] $=-1.0$. The dominant sources of uncertainty were attributed to $\alpha$-element abundance, the mixing length parameter, and the high-temperature opacities, all of which are increasingly important at higher metallicity. We stress their main result is that from a purely theoretical perspective, there is more scope for uncertainty in the metal-rich models. 
We also note the work by \citet{2011PASP..123..879T} who used the empirical brightness difference between
the LF bump and the point on the main sequence that is at the same
colour as the bump. Their models also underestimated the magnitude of the LF bump unless they employed an initial 
He mass fraction Y $=0.2$ which is lower than the Big Bang Nucleosynthesis value.

\subsection{Comparison Across the Globular Cluster Metallicity Distribution} \label{CAGCMD}
\begin{figure*}
 \includegraphics{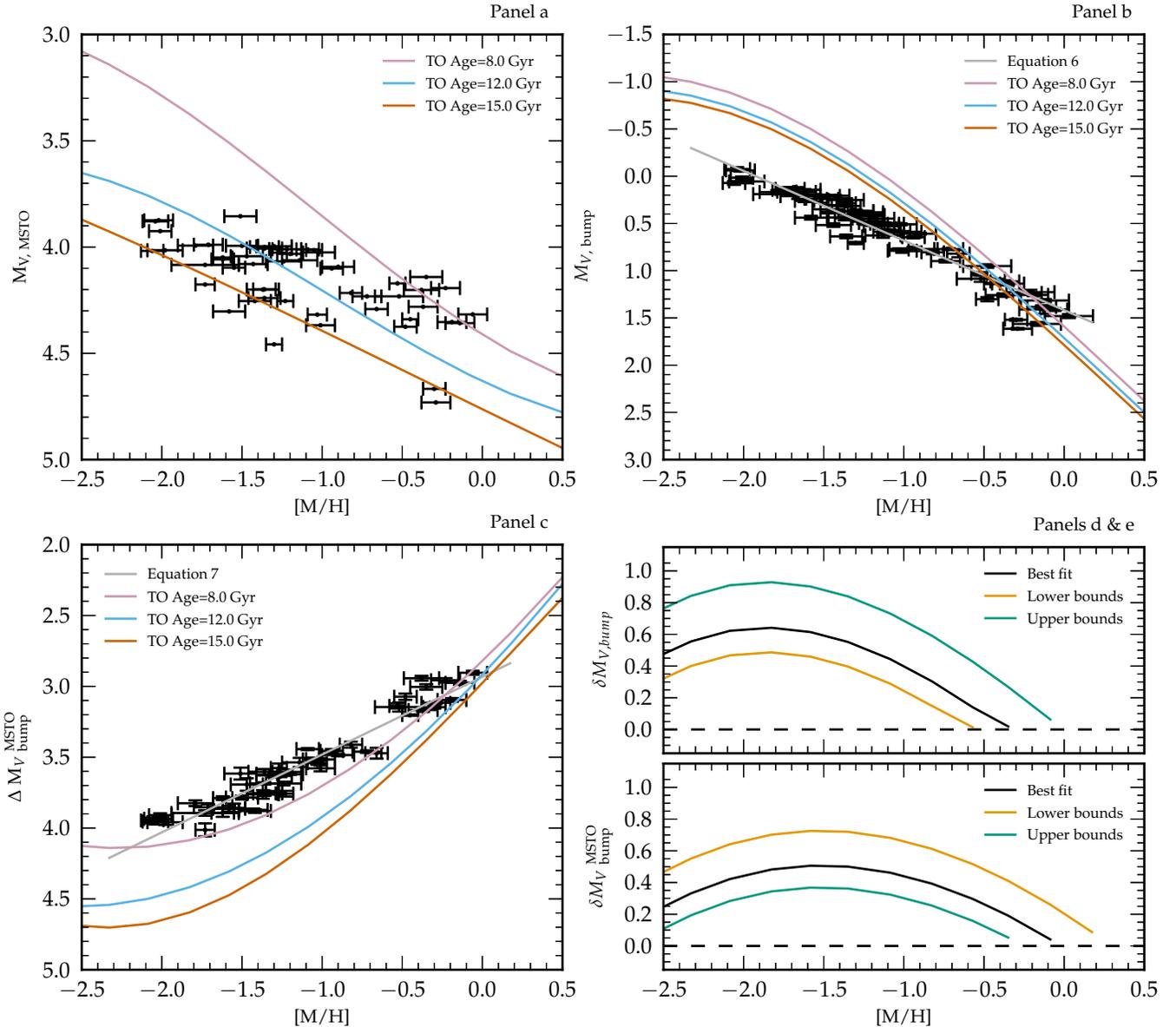}
\caption{Panel a: MSTO magnitudes for 55 GCs (black circles with associated uncertainties, \citealt{2013ApJ...766...77N})  as a
function of [M/H]. We include predictions from our stellar models assuming a MSTO age of 8 Gyr (lavender), 12 Gyr (blue) and 15 Gyr (vermilion).  
Panel b:  The empirically determined location of the LF bump in 72 GCs as a
function of [M/H]. The data are from the sources described in panel a and retain their symbols. The grey curve in this panel is the line of best fit to the data determined by \citet{2013ApJ...766...77N} and described by Equation 6. 
Panel c: The distance independent parameter $\Delta M_V ~^{\rm{MSTO}}_{\rm{bump}}$ as a function of [M/H]. The data are from the sources described in panel a and retain their symbols. The grey curve in this panel is the line of best fit to the data determined by \citet{2013ApJ...766...77N} and described by Equation 7.  Panel d: 
The Euclidean distance between the LF bump magnitude determined from the 12 Gyr isochrone and respective  best, lower bound and upper bound fits to the observational data in panel b.  
Panel e: The Euclidean distance between the $\Delta M_V ~^{\rm{MSTO}}_{\rm{bump}}$ parameter determined from the 12 Gyr isochrone and respective  best, lower bound and upper bound fits to the observational data in panel c.}
 \label{fig:nataf_1}
\end{figure*}

\begin{table}
%\begin{center}
\centering
\begin{tabular}{ccccc} 

  \hline \hline 
Z & [M/H] &  Mass ($\Mo$)  \\
\hline 
0.02	    &	0.17	    &	1.017	 \\
0.011	&	-0.09	&	0.956\\
0.006	&	-0.35	&   	0.900  	\\
0.0036	&	-0.57	&	0.854 \\
0.002	&	-0.82	&	0.828	 \\
0.0011	&	-1.09 	&	0.812	 \\
0.0006	&	-1.35	&	0.805	 \\
0.00035	&	-1.58 	&	0.800	 \\
0.0002	&	-1.83	&	0.800	 \\
0.00011	&	-2.09	&	0.800	\\
0.000063	&	-2.33	&	0.800	 \\
\hline

\end{tabular}
\caption{The metallicities of the grid of models used to generate isochrones in Figures \ref{fig:nataf_1} and \ref{fig:DMtest}. The third column indicates the initial mass of the star representative of the stellar structures populating the RGB at an age of 12Gyr.} 
\label{tab:natafmodels}
\end{table}
\subsubsection{Results and Discussion}
In Figure \ref{fig:nataf_1} we plot the magnitude of two key evolutionary features as a function of metallicity. 
Matching the MSTO luminosity (panel \ref{fig:nataf_1}a) and LF bump luminosity (panel \ref{fig:nataf_1}b) serve as initial tests for our models. To remove the effects of distance and reddening uncertainties, we also compare the magnitude difference between the locations of the MSTO and LF bump  (panel \ref{fig:nataf_1}c). In these three panels the observational data are plotted with black circles and taken from \citet{2013ApJ...766...77N} who performed statistical analyses on HST photometry. 
Their work yielded the MSTO magnitudes for 55 clusters to which we compare our models in  panels \ref{fig:nataf_1}a and \ref{fig:nataf_1}c. These 55 clusters are a subset of the 72 systems shown in \ref{fig:nataf_1}b for which they determined the LF bump magnitudes.

We have constructed a grid of models that vary in mass and metallicity to coincide with MSTO ages of 8 Gyr (lavender curve), 12 Gyr (blue curve) and 15 Gyr (vermilion curve). We fit third order polynomials through the quantities of interest to interpolate across the entire globular cluster metallicity range.  The metallicity spacing of our grid is listed in  Table \ref{tab:natafmodels} and is the same for each isochrone.  The 12 Gyr isochrone will serve as somewhat of a fiducial track in our analysis and in Table \ref{tab:natafmodels} we indicate the corresponding mass at each metallicity for this case. 
In all models we assume an initial hydrogen abundance of 
$0.75$ and we assume a solar scaled ASG09 metallicity mixture with no $\alpha$-element enhancement for the initial composition and opacity mixtures. 

To quantify the discrepancy between theory and observation at the LF bump, we compare our models with the lines of best fit to the data in panels \ref{fig:nataf_1}b and  \ref{fig:nataf_1}c (i.e., equations 1 and 2 in \citealt{2013ApJ...766...77N})
\begin{equation}
M_{\rm{V, bump}} = 0.600 + 0.737(\rm{[M/H]}+1.110)
\end{equation} 
\begin{equation}
\Delta M_V ~^{\rm{MSTO}}_{\rm{bump}} = 3.565 - 0.549(\rm{[M/H]}+1.152).
\end{equation}
These fits are translated to give upper and lower envelopes to the data in order to provide an indicative uncertainty at each metallicity. The discrepancy between each of these three fits and the 12 Gyr isochrone can be found in Panel \ref{fig:nataf_1}d, (top) for the direct comparison method and in panel \ref{fig:nataf_1}e (bottom) for the distance independent parameter. 

In panel \ref{fig:nataf_1}a  we run into the familiar result (see also figure 2 of Cassisi et al. 2011) that 15 Gyr isochrones are required to match the MSTO in many GCs. It is expected that  given the age constraints of 
Big Bang Cosmology and the cosmic microwave background \citep{2014A&A...571A...1P} that the 12 Gyr isochrone should provide an upper age limit to the clusters. The data demonstrates a clear spread in age at each metalliciy with a collection of clusters centred around [M/H] $\approx -0.3$ dex well modelled by the 8 Gyr isochrone.

The stellar models also tend to predict LF bump magnitudes that are too bright compared to the observational data. 
There is some agreement at the metal rich end of the spectrum above a metallicity of [M/H]$=-0.8$.
At the lowest metallicites  the models differ significantly -- by up to 0.8 magnitudes for some clusters (see panel \ref{fig:nataf_1}d). The fact that our models fail to reproduce the slope of the observational data suggests that we need to give careful consideration to the internal stellar processes and how they operate across different metallicities.

Similarly, we find some agreement for the distance-independent $\Delta M_V{}^{\rm{MSTO}}_{\rm{bump}}$ parameter at the highest metallicities. The level of discrepancy outlined in panel \ref{fig:nataf_1}e is consistent with a roughly constant offset of $\delta M_V \approx 0.4$. This is slightly higher than the value of $\delta M_V \approx 0.2$ found by \citet{2011A&A...527A..59C} but their models better reflect GC abundances (see below for discussions on \el{4}{He} and $\alpha$-elements).     
It is essential to point out the `better' agreement implied by the 8 Gyr isochrone.
This track significantly overestimates the MSTO luminosity and predicts an incorrect age in the low-metallicity clusters but in doing so, better matches the $\Delta M_V ~^{\rm{MSTO}}_{\rm{bump}}$ value.

\subsubsection{The Distance Modulus and Metallicity Scale}
\begin{figure}
 \includegraphics{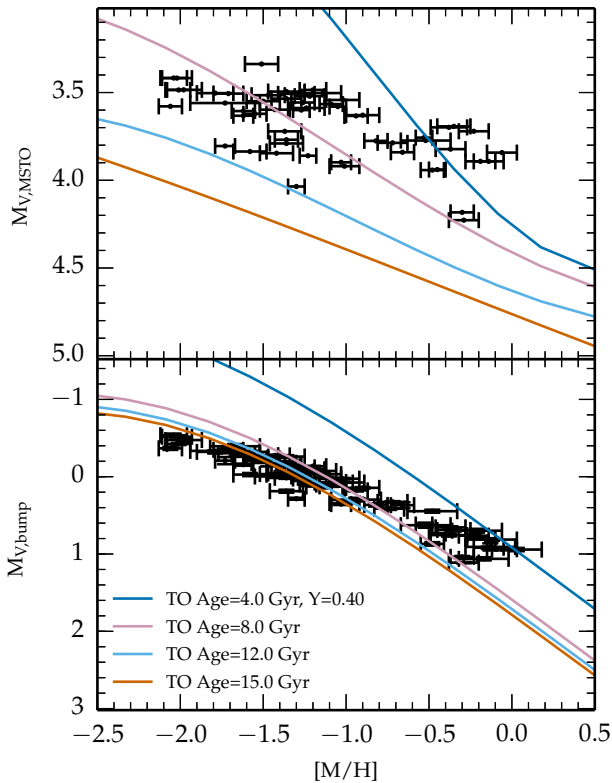}
\caption{Same as Figures \ref{fig:nataf_1}a and \ref{fig:nataf_1}b except assuming 3\% increase to the distance modulus for each cluster. Note the additional results for age$=4.0$ Gyr and Y$=0.40$.}
 \label{fig:DMtest}
\end{figure}

Our results, irrespective of how the comparison is carried out, imply that the stellar models do not correctly reproduce the stellar structure at low metallicity.
The MSTO and LF bump magnitudes are underestimated (too bright) in our models and we do not match the magnitude difference between these two evolutionary indicators. 
Naively, if we are unable to match the earlier evolutionary events then we have no right to expect to match the later events, but it is possible to shift the magnitude of the LF bump (essentially) independently of the MSTO.  The LF bump is of course dependent on the depth of FDU:  if FDU is deeper, then the hydrogen-burning shell will encounter the composition discontinuity at a lower luminosity. Improvements to stellar models can therefore be attained through either refinement of the microphysics or changes in the treatment of mixing. 

Before we investigate the role of the physics included in the stellar models, we first determine whether uncertainties in the observational data can help reduce the disagreement with theoretical predictions. One of our key questions pertains to the usefulness of direct comparisons and the uncertainty surrounding the distance modulus. A range of methods are used to determine the distance to GCs. The distance moduli listed in the \citet{1996AJ....112.1487H} catalogue (2010 edition) are predominately determined from calibrating the mean V magnitude of the horizontal branch. 
\citet{1996AJ....112.1487H} cites an uncertainty in determining this mean magnitude of at least $\pm 0.1 \ M_V$. There are also unquantified difficulties in the magnitude-metallicity relation used for calibration \citep{2000AJ....119.1398D}. 
In the case of  NGC 6397, the blue horizontal branch introduces much uncertainty for this method.  Its distance modulus was derived  by fitting isochrones to the MSTO \citep{1987AJ.....94..917A}. In fact most of these techniques used to determine the distance modulus require some form of calibration from isochrones. 
%This may suggest that the discrepancy in the direct comparison method may reflect systematic modelling differences between MONSTAR and those codes used to generate the isochrones.

Figure \ref{fig:DMtest} demonstrates how a systematic offset in the distance modulus determination would help to reconcile
models with the observational data. 
We have reproduced panels \ref{fig:nataf_1}a and \ref{fig:nataf_1}b but multiplied the distance modulus for each cluster by an \textit{ad hoc} factor of 1.03. 
The 12 Gyr MSTO models now matches the upper extent of the MSTO magnitudes across all metallicities.
Directly matching the LF bump with our isochrones requires that some metal-rich clusters possess a \el{4}{He} content of up to Y=0.4, however, this isochrone is unable to simultaneously provide a lower limit to the MSTO magnitude in these systems. Further changes to the input physics  are seemingly necessary to account for the systems with the brightest LF bumps. In fact, 4 Gyr is probably not a realistic age estimate for a GC given their metallicity. 
Nevertheless, it is  somewhat pleasing that parameters within known observational limits can reproduce \textit{most} of the data; 
a more reasonable state of affairs than requiring stars with ages greater than the Hubble Time.  
The uncertainties introduced by the absence of boutique modelling (where all known cluster parameters are considered)
and self-consistent determination of the distance modulus (where fitting is done with the code generating the models) may contribute to the need for this factor 1.03.

The chosen metallicity scale is also a source of uncertainty and has been discussed in previous studies \citep{2003A&A...410..553R, 2010ApJ...712..527D, 2011A&A...527A..59C}. \citet{2010ApJ...712..527D}, in particular, find that by adopting two different sets of independently determined metallicites for their GC sample, their conclusions remain unchanged. The models still underestimate the bump magnitude. 
We note that a systematic offset of about 0.5 dex in [M/H] would help reduce \textit{some} of the  inconsistency between theory and observation but there is currently no evidence for advocating this. 
Changes to the distance modulus and metallicity scale are equivalent to applying luminosity translations in the HR diagram. 
So whilst `corrections' to the observational data may assist with the direct comparison methods they do not help with the distance independent comparisons. The $\Delta M_V ~^{\rm{MSTO}}_{\rm{bump}}$ parameter is set by the stellar physics and essentially a constant offset at low metallicity.

\section{Investigation of the Stellar Physics}
The choice of stellar physics has some role to play in reconciling the differences between theoretical and observational determinations of the MSTO and LF bump magnitudes and, in particular, reducing the luminosity difference between these events. In the following sections we quantify how choices in the modelling affect the bump magnitude and the $\Delta M_V ~^{\rm{MSTO}}_{\rm{bump}}$ parameter.
Where possible, we generate new isochrones to illustrate the effect of the stellar physics on our parameters of interest, but
in some cases we compare individual stellar tracks because it makes the analysis simpler. We note that because of the short lifetimes of the SGB and RGB, single evolution tracks coincide almost exactly with isochrones. Our models that turn off the main sequence at an age of 12 Gyr will serve as an appropriate proxy for the isochrone and once the effects of the  microphysics are quantified, they can later be incorporated in the grids of models that generate the isochrones.

\subsection{Mass and \el{4}{He}-Enhancement}
\begin{figure}
 \includegraphics{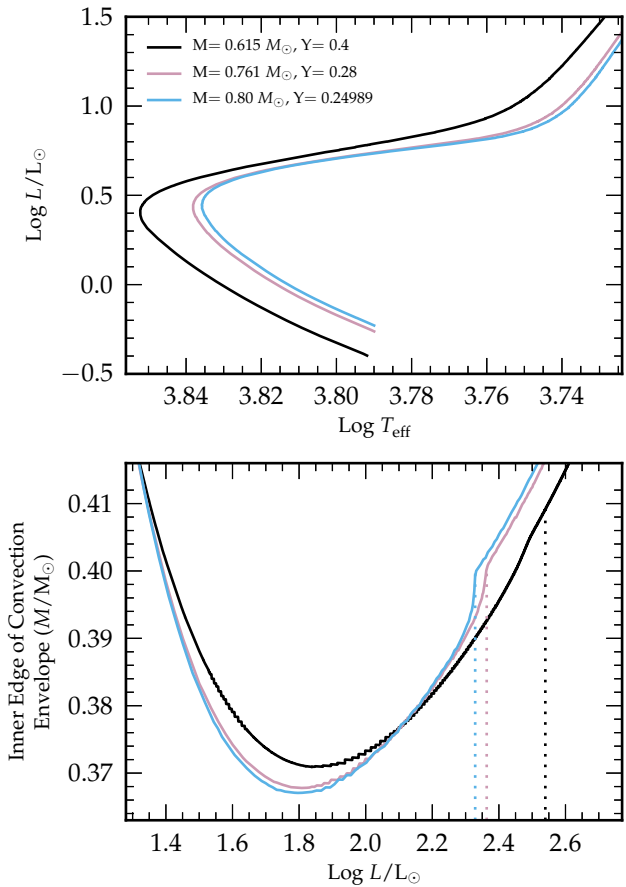}
\caption{Top panel: stellar tracks for stars representative of GC PIE populations (see definition in text) at a metallicity of Z=0.00011. Mass and helium abundances are as specified in the legend. Each model reaches the MSTO with an age of 12 Gyr. Bottom panel: penetration of the convective envelope as a function of luminosity for the respective models. Dotted vertical lines mark the bump location in each case.}
 \label{fig:massHe4}
\end{figure}

The existence of multiple populations in GCs is well documented (see \citealt{2012A&ARv..20...50G} and references therein). 
The constituent stars divide into chemically distinct groups that are best explained by explicit stellar generations. The most commonly proposed scenario is that a fraction of first generation stars pollute the environment from which the subsequent generation(s) form (which we herein refer to as second generation). More-massive stars in the first generation, having 
experienced the advanced phases of stellar nucleosynthesis, burn hydrogen via the CNO cycles. The degree of mixing between first-generation ejecta and pristine cluster gas determines whether the second generations are `extreme' population stars (sometimes formed from pure ejecta) or belong to the `intermediate' population that give rise to the observed anticorrelations (mixed ejecta and cluster gas, see  \citealt{2009A&A...505..117C} for discussions on [P]rimordial, [I]ntermidiate and [E]xtreme populations [PIE]). The extreme population is not evident in every GC, however, it is predicted that all second generation stars are necessarily enhanced in helium compared to their primordial counterparts.  
 A spread in \el{4}{He} abundance also straightforwardly explains the horizontal branch morphology in many clusters \citealt{2008MNRAS.390..693D,2014MNRAS.437.1609M} and is consistent with observed split main sequences \citep{2007ApJ...661L..53P,2012ApJ...754L..34M}.    

The effect of \el{4}{He} on the stellar models is well documented in the literature --
originally in a series of papers by Iben \citep{1968Natur.220..143I, 1968ApJ...153..101I} and more recently by \citet{2013MmSAI..84...91C}.  \citet{2006ApJ...645.1131S} and \citet{2011A&A...527A..59C} place an upper limit of $\Delta X(\el{4}{He}) =0.05$ on the width of the main sequence for those clusters that do not display split evolutionary tracks. This in turn decreases the TO age by no more than $\approx 0.5$ Gyr. In clusters such as $\omega$ Cen, a spread in metallicity and split main-sequences imply a wider span in age, however, these complex systems are in the minority.  In most cases, GCs are dominated by second generation stars with an inferred separation in age of at most a few hundred Myr \citep{2010A&A...516A..55C}. 
Stellar models representative of these stars only further exacerbate the discrepancies demonstrated in Figure \ref{fig:nataf_1}.
In Figure \ref{fig:massHe4} we plot the HR-diagram (top panel) and penetration of the convective envelope (bottom panel) for three stellar models representative of the PIE populations at a metallicity of
Z=0.00011. Each model, with its combination of initial mass and  \el{4}{He} content leaves the main sequence with an age close to 12 Gyr.
Our models are by design the same age and metallicity so there are minor differences in their MSTO luminosity but they differ in temperature. In the bottom panel the luminosity of the bump for each model is marked by the dotted vertical lines. 
The impact of enhanced helium on the envelope opacity reduces the extent of FDU, and delays the onset of the bump ($\approx$ 0.2 Log L/$\Lo$). The second generation models (see also the Y=0.40 isochrone in Figure \ref{fig:DMtest} and \citealt{2006ApJ...645.1131S}) are brighter, contrary to the need for fainter MSTO and LF bump magnitudes demonstrated in Figure \ref{fig:nataf_1}. It follows that,  as per \citet{2011PASP..123..879T}, primordial \el{4}{He} abundances below the accepted Big Bang nucleosynthesis level would allow models to better reproduce observations.

\subsection{Opacity} 
\begin{figure}
 \includegraphics{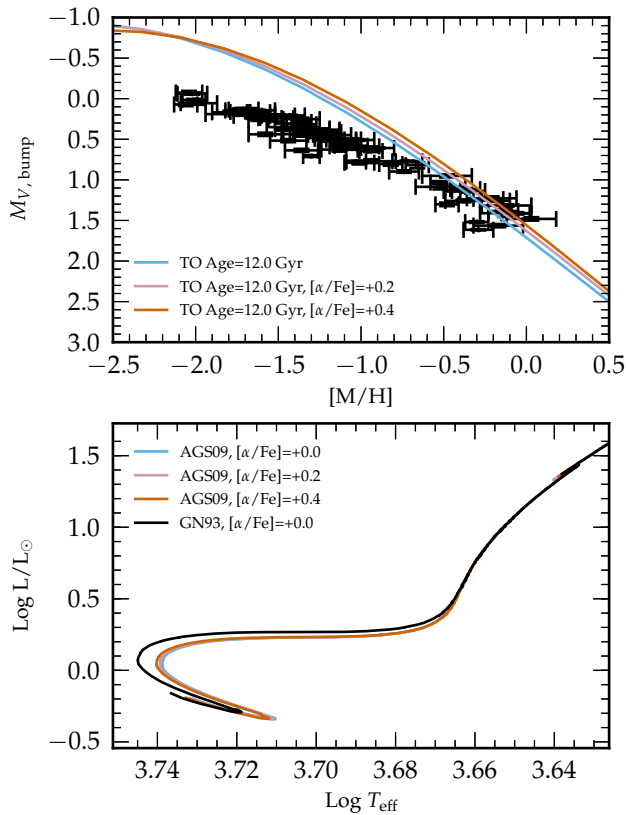}
\caption{Top panel: The predicted bump magnitude for 12 Gyr isochrones with various levels of $\alpha$-element enhancement. Bottom panel: HR diagram for the 12 Gyr isochrone with AGS09 solar scaled oxygen abundance but different selections of $\alpha$-element enhancement in the heavy mixture opacity tables. }
 \label{fig:opacal}
\end{figure}

The models presented thus far do not account for the effect of  $\alpha$-element enhancement on the evolution/isochrones. The inclusion of $\alpha$-elements has a two-fold effect. First, it increases the amount of available CNO material which impacts upon the nuclear burning. Second, with a larger reservoir of metals, there is a contribution to the opacity that is predominantly manifested as a reduction of the effective temperature. As a function of [M/H], the inclusion of  $\alpha$-element enhancement results in only minor changes to the MSTO luminosity.

In the top panel of Figure \ref{fig:opacal}  we plot the bump magnitude as a function of [M/H] determined from 12 Gyr isochrones with  [$\alpha$/Fe]$=0.0$ (blue curve), [$\alpha$/Fe]$=0.2$ (vermilion curve) and [$\alpha$/Fe]$=0.4$ (lavender curve).
The heavy element mixtures for the respective opacity tables can be found in the appendix. $\alpha$-element enhancement shifts the bump to brighter magnitudes in the more metal-rich models. This effect is on account of the increased O available in the nuclear burning chains. In the scaled-solar case  $\alpha$-element enhancement of [$\alpha$/Fe]$=0.4$ shifts the bump by approximately 0.15 mag.

The bottom panel of \ref{fig:opacal} illustrates the effect of the different opacity tables on the individual stellar models.     
We have calculated evolutionary tracks based on a star with metallicity Z=0.02 and TO age of 12 Gyr used in generating our isochrones. 
In these calculations we employ $\alpha$-enhanced opacity tables but do not change the composition of the models.  
The CNO abundances in our energy generation network are kept at the ASG09 scaled-solar values, which removes the impact of increased CNO material on the bump location. The opacity tables used are described in the top panel and detailed in the appendix.
We also include a calculation (black curve) with the GN93 scaled-solar composition and find variations of up to 0.04 Log L/$\Lo$ are possible in the bump location.

\subsection{Mixing Length Parameter}
The free parameter in the MLT is also a source of uncertainty in the stellar models. $\alpha_{MLT}$ is calibrated to the Sun, a star halfway through its main-sequence lifetime which possesses a shallow convective envelope (in mass). This value is then applied across the stellar mass and metallicity distribution and through all phases of evolution. Work by \citet{2015A&A...573A..89M} suggests that this should in fact vary with evolutionary status but the solar calibrated mixing length parameter does allow a majority of the model libraries e.g., BASTI \citep{2004ApJ...612..168P}, Padova \citep{2012MNRAS.427..127B} and Victoria \citep{2006ApJS..162..375V} to match the effective temperature of  RGB stars in GCs (see also the work by \citealt{2015arXiv150304582S}).
The value of $\alpha_{MLT}$ adopted will significantly shift the model's position in the HR diagram because the convective efficiency will change the surface temperature.  The effect is greatest on the RGB where the stars possess deep convective envelopes. 
Although not illustrated here, we find that the depth of FDU is not altered significantly by this parameter (see also \citealt{1991A&A...244...95A}) and we find variation of approximately 0.05 mag between $\alpha_{MLT}=1.60$ and $\alpha_{MLT}=1.85$.

\subsection{Equation of State}
\begin{figure*}
 \includegraphics{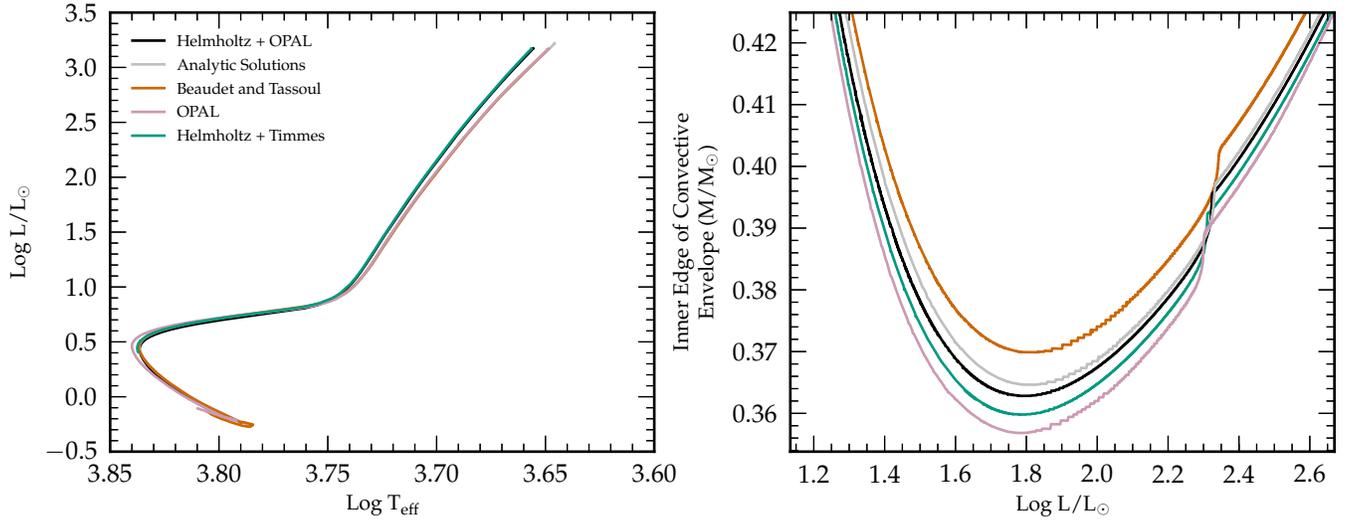}
\caption{Left panel: Evolution through the HR-diagram of the stellar model with metallicity Z=0.00011. Each curve corresponds to a different choice for the EOS. Right panel: Penetration of the convective envelope for each EOS choice.}
 \label{fig:eos}
\end{figure*}
MONSTAR typically employs the fitting formula by  \citet{1971A&A....13..209B} whilst partially-ionized regions are treated with the Saha equation as described by \citet{1965ZA.....62..221B}.  In the event of convergence issues, a computationally more expensive analytic solution can be calculated but this method still requires the numerical evaluation of the Fermi-Dirac integrals. To aid in this investigation we have added the OPAL EOS tables from \citet{2002ApJ...576.1064R}. We also include models from MONSTAR incorporating
the Timmes EOS \citep{1999ApJS..125..277T} and the Helmholtz equation of state \citep{2000ApJS..126..501T} which is the tabulated form of the Timmes EOS \citep{2014ApJ...784...56C}. The Timmes EOS has allowances for ``simple ionisation" that includes a simple two-level hydrogen-like atom model for ionization. In cases where equations of state are blended, a linear transition occurs over the temperature range T $=2 - 2.5$ MK. 
%In order to ensure a consistent comparison with work by \citet{2014ApJ...784...56C}, the GN93 opacity tables are used for the models presented in Figure \ref{fig:eos}. 

We have calculated the evolution of a star with mass M$=0.8 \ \Mo$ and metallicity Z=0.00011 using 
five different EOS combinations (Figure \ref{fig:eos}). Employing the Helmholtz EOS in the high temperature regime produces a systematically higher \Teff\ at a given luminosity compared with other choices for the EOS.  The depth of FDU for each combination is plotted in the adjacent panel. The \citet{1971A&A....13..209B} fitting formula, which is the default choice in MONSTAR, provides the shallowest penetration of the convective envelope of all five EOS combinations. This choice of EOS therefore yields the greatest discrepancy with the empirically determined LF bump magnitude.     
The deepest penetration of the convective envelope is achieved by using the OPAL EOS and leads  to approximately 0.1 magnitude better agreement with the empirical data. 

\subsection{Overshoot}
\begin{figure*}
 \includegraphics{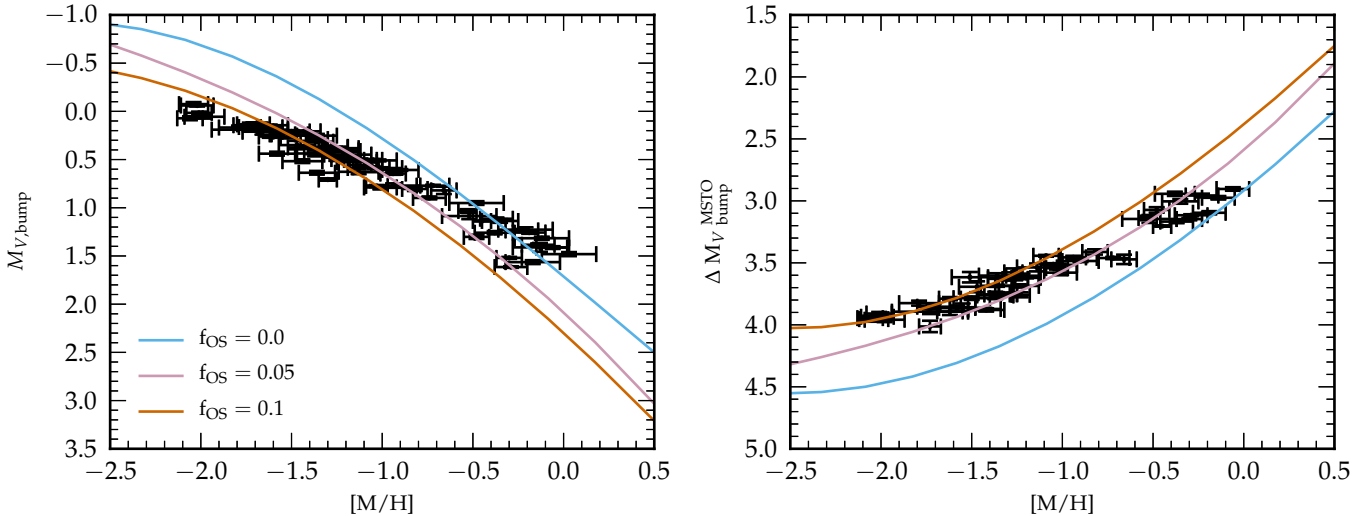}
\caption{Left panel: LF bump magnitudes as a function of metallicity from the observations specified in Figure \ref{fig:nataf_1} (black circles). The coloured curves denote our 12 Gyr isochrone with various levels of overshoot at the base of the convective envelope. Right panel: $\Delta M_V{}^{\rm{MSTO}}_{\rm{bump}}$ parameter for the observational data specified in Figure \ref{fig:nataf_1} (black circles) and for our 12 Gyr isochrone with various levels of overshoot (coloured curves) at the base of the convective envelope. }
 \label{fig:os}
\end{figure*}

Once it had been identified that stellar models overestimate the bump luminosity \citep{1985ApJ...299..674K, 1990A&A...238...95F},   
\citet{1991A&A...244...95A} suggested that overshooting by $0.7-1.0$ pressure scale heights ($H_{\rm{P}}$) at the base 
of the convective envelope would provide the required shift in magnitude to reconcile theory with observations. 
Using the formalism outlined in \S \ref{modov} we find that values of $f_{os}=0.05-0.1$ significantly improve the agreement with the observations -- both with the direct comparison of the bump magnitude and with the  $M_V ~^{\rm{MSTO}}_{\rm{bump}}$ parameter. In Figure \ref{fig:os} we plot the 12 Gyr isochrone with no overshoot (blue curve), $f_{os}=0.05$ (lavender curve) and  $f_{os}=0.1$ (vermilion curve). We do not expect that all GCs are 12 Gyr old, but if these tracks can provide appropriate extrema to the data in Figure \ref{fig:os} then a spread in age and overshoot efficiency may allow models to reproduce the observations.    

The isochrones reproduce the gradient of the $M_V ~^{\rm{MSTO}}_{\rm{bump}}$ parameter, 
but do not match the bump magnitudes.  The isochrones still do not provide an upper limit to the some of the most metal-poor clusters but do satisfactorily explain intermediate-metallicity and metal-rich clusters.   
This may hint that at the need for an overshoot formalism that is a function of metallicity such that:
\begin{itemize}
\item for [M/H]  $> 0 $ no overshoot is required;
\item between  $ -0.8 \lesssim $ [M/H] $ \lesssim 0.0 $ overshoot increases to $f_{os} \approx 0.05$;
\item between $-1.3 \lesssim$  [M/H] $\lesssim -0.8$ overshoot increases to $f_{os} \approx 0.1$;
\item for [M/H] $< -1.3 $ overshoot $f_{os} > 0.1$ is required.
\end{itemize}

It is interesting to note that our models with overshoot cannot simultaneously provide a lower limit to the absolute magnitude of the bump and the $M_V ~^{\rm{MSTO}}_{\rm{bump}}$ parameter. In the left panel of Figure \ref{fig:os} the most efficient form of overshoot we have modelled predicts a bump location that is approximately 0.3 magnitudes too bright in the most metal-poor clusters. 
If we include more efficient overshoot such that our curve encompasses the bump magnitude of these systems, it will necessary reduce the separation between the 
MSTO and LF bump. This is in spite of the fact the vermilion curve ($f_{os}=0.1$) currently forms a satisfactory lower envelope.    
The discrepancy between the two panels implies that the models predict a MSTO magnitude that is too bright for a 12 Gyr isochrone -- an outcome of different input physics. Such a result reflects the fact 15 Gyr isochrones are required to account for the MSTO luminosities in Figure \ref{fig:nataf_1}a.

\subsection{Other Input Physics}

Determinations of the key hydrogen burning reaction rates have changed little since \citet{1983ARAA..21..165H} save for the significant revision of the \el{14}{N}(p, $\gamma$)\el{15}{O} rate which we have discussed in \S \ref{modov} and use in the current calculations. One possible source of improvement lies with the recent measurement of the iron opacity at temperatures pertaining to solar interiors. \citet{ironopac} find that the measured iron opacity is much higher (30-400\%) than previously predicted. In their series of tests \citet{1993ApJ...414..580S} demonstrate the significant role that the high temperature opacities have on the luminosity of the MSTO. An increase in Fe opacity will help reduce the MSTO luminosity. Only a small improvement is likely because the change is wavelength dependent and Fe is just one of many elements that contribute to the opacity. 

The inclusion of additional physical processes is an obvious avenue of pursuit  with the role of atomic diffusion especially a contentious issue (see \citealt{2013A&A...555A..31G} and references therein). If the process is not inhibited then atomic diffusion will reduce the MS lifetime and hence the turn-off luminosity \citep{2001ApJ...562..521C}. Observational evidence however \citep{2001A&A...369...87G} suggests that the process is indeed inhibited. \citet{2011A&A...527A..59C} state that the overall effect from including diffusion on $M_V ~^{\rm{MSTO}}_{\rm{bump}}$ is a reduction of approximately 0.05 mag. The role of rotation and magnetic fields are also uncertain but their inclusion will have consequences for the entire evolution.  \citet{Chaname2005} and \citet{2006A&A...453..261P} demonstrate how the treatment of angular momentum impacts upon the stellar models.  In their comparison of different transport prescriptions, \citet{2006A&A...453..261P} found a negligible difference in MSTO luminosities and the depth of FDU. However, each of the rotating models still predicted a higher LF bump luminosity than in the non-rotating models. The higher-mass higher-metallicity ($M \geq 1.25 \ \Mo$, $Z=$Z$_{\odot}$) models by \citet{2010A&A...522A..10C}, on the other hand, demonstrate a significantly lower bump luminosity in their rotating models compared to their non-rotating models.

\section{Second Analysis of the Spectroscopic Data: Thermohaline Mixing}
 \label{testthm}

\begin{figure}
 \includegraphics{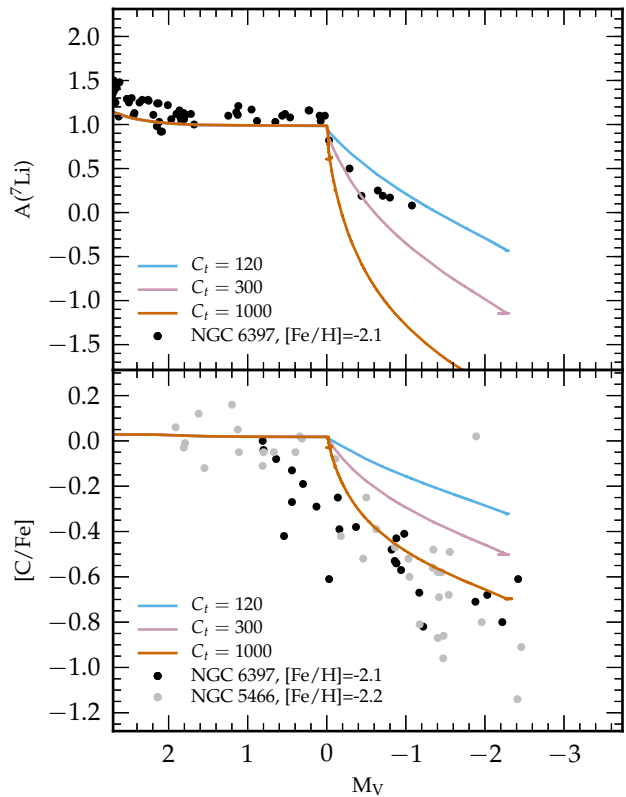}
\caption{Top panel: A(Li) as a function of magnitude for NGC 6397 ([Fe/H] $=-2.1$).  We include appropriate models with three variations of the thermohaline mixing parameter $C_t$. Blue curves denote $C_t=120$, lavender denote $C_t=300$ and vermilion curves denote $C_t=1000$. Bottom Panel: Panel b: [C/Fe] as a function of luminosity in  NGC 6397 (black circles) and NGC 5466 ([Fe/H] $=-2.2$, grey circles). Model data are as described for the above panel. Sources for the observational data (in both panels) are described in the main text.}
 \label{fig:10a}
\end{figure}

\begin{figure*}
 \includegraphics[scale=1]{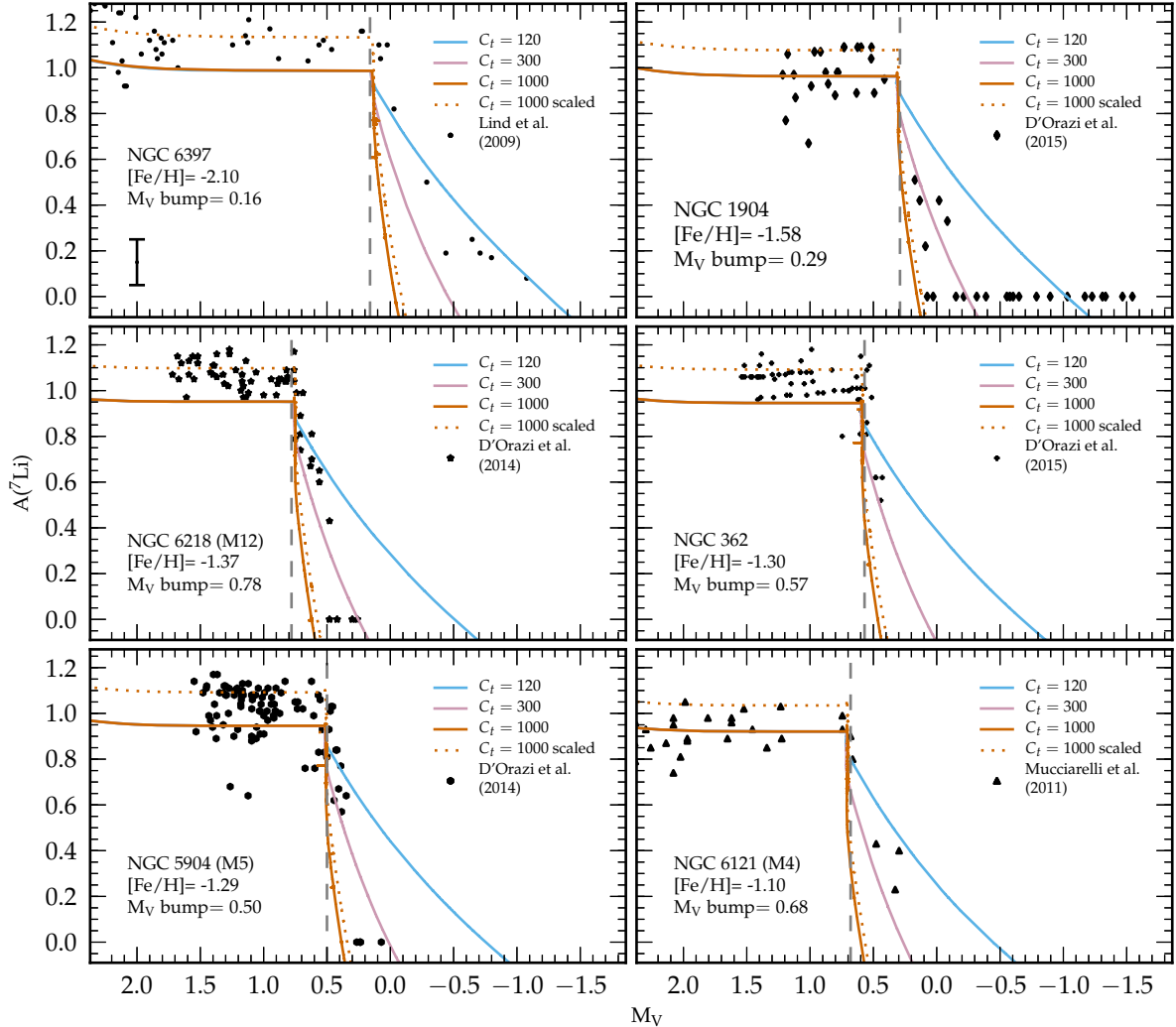}
\caption{A(Li)as a function of magnitude for six of the clusters described in Table \ref{tab:clusters}. For each panel we calculate stellar models that best represent the metallicity of the respective clusters. We include three variations of the thermohaline mixing parameter $C_t$ as described in the legend. We also indicate the location of the bump magnitude with the vertical dashed line and the sources from which the observational data are taken. The error bar in the top-left panel indicates the typical uncertainty in the data.}
 \label{fig:10b}
\end{figure*}

 In this section, we compare observations and theoretical predictions of the evolution of two species that are involved in different channels of hydrogen burning: the fragile lithium and the more robust carbon.  We have run a grid of models rather than conducting boutique modelling of each cluster.  All models were run with initial X(H)$=0.75$ and X(Li)$=9.39 \times 10^{-10}$ with an assumed TO age of 12 Gyr. Overshoot (with $f_{os}=0.075$) has been included  and ensures that the  $\Delta M_V{}^{\rm{MSTO}}_{\rm{bump}}$ parameter is consistent  with the value determined from the respective CMDs (but does not necessarily reproduce the bump magnitude). 
In order to test the efficacy of the thermohaline process, we have applied artificial offsets in magnitude to the models so that the beginning of extra mixing coincides with the photometrically derived LF bump. 

In the top panel of Figure \ref{fig:10a} we once again plot lithium observations of NGC 6397 (Lind et al. 2009, black circles) but include models with three different choices for the free parameter in the diffusive-thermohaline mixing theory ($C_t$). Although $C_t=1000$ has been the preferred value in the literature, matching abundances in many stellar environments, values  between  $C_t=120-300$ give a far better match to Li in this system. The same models are included in the bottom panel of Figure \ref{fig:10a} 
in our comparison to [C/Fe] data in NGC 6397 and NGC 5466. [C/Fe] data for NGC 6397 were 
taken from \citet{1990ApJ...359..307B} but determined by assuming that there is no oxygen over-abundance in that system. 
[C/Fe] in NGC 5466 is taken from \citet{2010AJ....140.1119S}. We expect that the models are representative of both clusters despite the small difference in metallicity.   NGC 5466 was one of the three metal-poor GCs studied by \citet{2012ApJ...749..128A} in their investigation of thermohaline mixing. Of the three, this was the only system in which thermohaline mixing appeared to reproduce the depletion of [C/Fe] along the RGB.  NGC 6397 seems to display the same ambiguity as its more massive counterparts:  it is difficult to identify the onset of extra mixing from the [C/Fe] data and one could argue that it begins well before the expected bump.  
[C/Fe] in NGC 5466, having been homogeneously analysed from a cluster of comparable size to NGC 6397, serves as the best analogue to the Li data. 

We find that a mixing efficiency of $C_t=1000$ depletes Li much too fast (top panel Figure \ref{fig:10a}) but is able to account for the depletion of [C/Fe] in both clusters (bottom panel Figure \ref{fig:10a}). In Figure \ref{fig:10b} we investigate lithium depletion the clusters listed in Table \ref{tab:clusters} with the exception of NGC 2808 where the data does not include stars experiencing extra mixing. In most cases our models provide a lower limit to the post-FDU Li abundances (on account that we have not tuned the initial abundances for each cluster). To guide the eye, we have artificially translated the $C_t=1000$ models to a higher post-FDU abundance (dotted vermilion curves) to indicate how depletion would proceed with boutique abundance choices of each system. In nearly all GCs the models with  $C_t=120$ and $C_t=300$  provide preferable fit to the Li data. The result is not entirely clear cut, especially in in the case of NGC 1904. However, we note that it is only the stars that have severely depleted Li that introduce some doubt. Those stars that are experiencing extra mixing (just after the bump) are better reproduced by the lower $C_t$ values thus these are our preferred values for reproducing the depletion of Li in GCs. In the bottom panel of Figure \ref{fig:10a} it is clear that with values of  $C_t=120$ and $C_t=300$ not enough carbon is processed to match the trend of [C/Fe]. We are unable to simultaneously account for [C/Fe] and A(Li) with the same free parameter. Matching both these abundances poses a challenge for any postulated mixing mechanism. The fragility of Li and robustness of C must be reproduced.

It may be that thermohaline mixing is not responsible for the surface composition changes of RGB stars. 
Objections to the mechanism are based on theoretical grounds and 3D hydrodynamical simulations \citep{2011ApJ...728L..29T, 2011ApJ...727L...8D, 2013ApJ...768...34B} that suggest the mixing is inefficient in the RGB regime. But we have described how the mechanism at work here can be considered a physically based phenomenological model.  Rather than mixing to a constant temperature or mass location from the hydrogen burning shell, material is transported to where \el{3}{He} burning creates an inversion in the $\mu$ profile. 
\citet{Church11092014} show how this location depends on the shell-burning conditions that change with RGB luminosity. In this extra mixing formalism we have varied the free parameter that controls the mixing rate. We note that \citet{2008ApJ...684..626D} have advocated that a deeper (than the $\mu$ inversion) but slower mixing may be required. It is not clear yet if this scenario could explain the abundance trends here. 
 
We have modelled thermohaline mixing as a diffusive process. Our evolution code employs a one-dimensional diffusion algorithm to model a process that is inherently three dimensional and advective.  Convection is characterised by streams (or plumes) of upward and downward travelling material. The majority of the nuclear processing in the extra mixing region occurs where material
turns over near the hydrogen-burning shell. A diffusive treatment of mixing assumes that the composition difference between convective elements at the same mass coordinate is negligible.  If the mixing speed in the extra mixing region is much slower than convection (as it generally is believed to be), then the turnover time is closer to the reaction timescale. We therefore expect 
a difference between the compositions of material approaching and returning  from the envelope at the same mass coordinate. 
It may in fact be necessary to develop a more realistic treatment of convection to better model the extra mixing process. 
Codes that calculate mixing via advective streams may prove to be enlightening \citep{1993MNRAS.263..817C, 2008A&A...490.1181S, 2009PASA...26..217C}.
\section{Conclusions}

We have utilised photometric and spectroscopic measurements of GCs to investigate the current state of red giant branch modelling. 
Observations at key evolutionary epochs, namely the MSTO, FDU and LF bump, provide a series of constraints for our stellar models. Li
determinations in seven GCs covering a factor of ten in metallicity have served as the key spectroscopic diagnostic. The surveys of NGC 6397, the most metal-poor cluster in our sample, and  M4, the most metal-rich, are of particular interest as they cover a luminosity range that extends from the subgiant branch to beyond the LF bump. Homogeneous Li data, spanning the key evolutionary indicators  has allowed us to confirm the long-suspected ambiguity associated with inferring the onset of RGB mixing from [C/Fe]. The many uncertainties in determining the [C/Fe] combine to give the appearance that mixing begins prior to the LF bump. In the seven clusters analysed here the onset of lithium depletion coincides with the photometric bump magnitude. 

We determined that the abundance changes predicted by FDU in the models are consistent with spectroscopic measurements, but we also found evidence that the luminosity at which mixing begins in the low-metallicity clusters is slightly overestimated. Data from a  larger sample of clusters and predictions from other stellar evolution codes will help confirm this behaviour.
 The luminosity that coincides with onset of extra mixing was significantly overestimated by our models, especially in the low-metallicity regime. By considering models of these two RGB mixing events we found that stellar codes better reproduce the structure of high metallicity stars. To further examine this hypothesis we turned to photometric data of the MSTO and LF bump in a large sample of GCs.

We directly compared the predicted MSTO and bump magnitudes as well as the parameter $\Delta M_V{}^{\rm{MSTO}}_{\rm{bump}}$ to the corresponding values measured from photometry. We reproduced two well known results:  i) that 15 Gyr isochrones are required to match the MSTO in some clusters and ii) that the models match the bump luminosity at the highest metallicities, but significantly overestimate it at low-metallicity. There is currently no reason to doubt the bolometric corrections at low-metallicity and there is no evidence to suggest that uncertainties in the metallicity scale are significant enough as to be responsible for the discrepancy. A systematic underestimation of the distance modulus by a factor of 1.03 would see 12 Gyr isochrones provide an upper limit to the ages of the GCs but would shift the models out of agreement with metal-rich clusters. A systematic error in the distance modulus would not explain the fact that the models do not reproduce the observationally determined  $\Delta M_V{}^{\rm{MSTO}}_{\rm{bump}}$ which is independent of  distance and reddening.

Changes to the stellar physics are the only means by which the theoretical value of $\Delta M_V{}^{\rm{MSTO}}_{\rm{bump}}$ can be made consistent with observations. Our tests demonstrated that reasonable variations to the stellar microphysics do not significantly alter the stellar structure. Our investigations 
covered composition and $\alpha$-element enhancement, opacity, convective efficiency and the EOS choice on RGB evolution. 
Of these,  updating MONSTAR to the most recent OPAL EOS yielded the greatest improvement ($\approx$ 0.1 mag compared to the $\gtrsim 0.4$ mag required).  
The inclusion of diffusive overshoot in the models is sufficient to 
reproduce the bump magnitude and  
$\Delta M_V{}^{\rm{MSTO}}_{\rm{bump}}$ at intermediate and high metallicity. In order to directly match the bump magnitude at low-metallicity the efficiency of overshoot needs to be increased, so much so that agreement with  $\Delta M_V{}^{\rm{MSTO}}_{\rm{bump}}$ is lost.  
The fact that our models with overshoot cannot simultaneously provide a limit (as a function of [M/H]) to the absolute magnitude of the bump and $\Delta M_V{}^{\rm{MSTO}}_{\rm{bump}}$ is most likely due to predicted MSTO magnitudes that are too bright for 12 Gyr isochrones.

The final test of the models focussed on their ability to reproduce the abundance changes associated with extra mixing. 
 We determined that the thermohaline mechanism could not simultaneously account for the reduction of [C/Fe] and A(Li) as a function of RGB luminosity. As it stands, current observations imply that at low metallicity Li and carbon begin mixing a different luminosities. This result is independent of models and seemingly impossible. In the six GCs examined, a free parameter between  $C_t=120-300$ provided a bounds to the Li data. This is in contrast to a value of $C_t=1000$ cited throughout the literature to match the carbon abundances in various stellar environments. Whilst this discrepancy is a challenge for thermohaline mixing, the constraints may prove difficult for other mechanisms also, as Li is likely to be destroyed very efficiently under the conditions required to deplete C. 
Finally we suggest that the diffusive mixing algorithm used in stellar modelling may not be adequate to follow elements such as Li that are very sensitive to the local thermodynamic history. In order to quantitatively match observations numerical codes may require a more realistic treatment of mixing.

\section*{Acknowledgements}
We warmly thank our anonymous referee whose comments significantly improved this manuscript.
The authors would like to thank Carolyn Doherty, Simon Campbell, Luca Casagrande and Warrick Ball for useful discussions. We also thank David Nataf and Alessio Mucciarelli for providing us with their data tables. GCA would like to thank the staff at MoCA for their help and support through the duration of his Ph.D. 
Part of the research leading to the presented results has received funding from the European Research Council under the European Community's Seventh Framework Programme (FP7/2007-2013) / ERC grant agreement no 338251 (StellarAges). VD is partially supported by an ARC grant (Super Science Fellow 2011).  RPC is supported by the Swedish Research Council (grants 2012-2254 and 2012-5807). RJS is a recipient of a Sofja Kovalevskaja Award from the Alexander von Humboldt Foundation.  by ARC DP1095368 and Discovery Grant
This work was supported in part by ARC DP1095368 and Discovery Grant DP120101815 (J. Lattanzio, P. I.) at Monash University,
Clayton, Australia.

\appendix
\section{Heavy Element Mixtures}

\begin{table*}
\centering
\label{tab:heavyel}
\begin{tabular}{cccccccccc} 

  \hline \hline \\
&& \multicolumn{2}{ c }{[$\alpha$/Fe]=0.0} &&
\multicolumn{2}{ c }{[$\alpha$/Fe]=0.2} &&
\multicolumn{2}{ c }{[$\alpha$/Fe]=0.4} \\
\cline{3-4}  \cline{6-7} \cline{9-10}\\
&&[Ni/Nz]&[i/Fe]&&[Ni/Nz]&[i/Fe]&&[Ni/Nz]&[i/Fe]\\
\hline \\
C	&&	2.58$\times 10^{-1}$		&	0.0		&&	1.95$\times 10^{-1}$		&	0.0	&&	1.41$\times 10^{-1}$		&	0.0\\
N	&&	6.49$\times 10^{-2}$		&	0.0		&&	4.90$\times 10^{-2}$		&	0.0	&&	3.53$\times 10^{-2}$		&	0.0\\
O	&&	4.71$\times 10^{-1}$		&	0.0		&&	5.63$\times 10^{-1}$		&	0.2	&&	6.43$\times 10^{-1}$		&	0.4\\
Ne	&&	8.17$\times 10^{-2}$		&	0.0		&&	6.17$\times 10^{-2}$		&	0.0	&&	4.45$\times 10^{-2}$		&	0.0\\
Na	&&	1.67$\times 10^{-3}$		&	0.0		&&	1.26$\times 10^{-3}$		&	0.0	&&	9.09$\times 10^{-4}$		&	0.0\\
Mg	&&	3.82$\times 10^{-2}$		&	0.0		&&	4.57$\times 10^{-2}$		&	0.2	&&	5.22$\times 10^{-2}$		&	0.4\\
Al	&&	2.71$\times 10^{-3}$		&	0.0		&&	2.05$\times 10^{-3}$		&	0.0	&&	1.47$\times 10^{-3}$		&	0.0\\
Si	&&	3.11$\times 10^{-2}$		&	0.0		&&	3.72$\times 10^{-2}$		&	0.2	&&	4.25$\times 10^{-2}$		&	0.4\\
P	&&	2.47$\times 10^{-4}$		&	0.0		&&	1.86$\times 10^{-4}$		&	0.0	&&	1.34$\times 10^{-4}$		&	0.0\\
S	&&	1.27$\times 10^{-2}$		&	0.0		&&	1.52$\times 10^{-2}$		&	0.2	&&	1.73$\times 10^{-2}$		&	0.4\\
Cl	&&	3.04$\times 10^{-4}$		&	0.0		&&	2.29$\times 10^{-4}$		&	0.0	&&	1.65$\times 10^{-4}$ 	&	0.0\\
Ar	&&	2.41$\times 10^{-3}$		&	0.0		&&	1.82$\times 10^{-3}$		&	0.0	&&	1.31$\times 10^{-3}$		&	0.0\\
K	&&	1.03$\times 10^{-4}$		&	0.0		&&	1.23$\times 10^{-4}$ 	&	0.2	&&	1.40$\times 10^{-4}$ 	&	0.4\\
Ca	&&	2.10$\times 10^{-3}$		&	0.0		&&	2.52$\times 10^{-3}$ 	&	0.2	&&	2.87$\times 10^{-3}$		&	0.4\\
Ti	&&	8.56$\times 10^{-5}$		&	0.0		&&	1.02$\times 10^{-4}$		&	0.2	&&	1.14$\times 10^{-4}$ 	&	0.4\\
Cr	&&	4.20$\times 10^{-4}$		&	0.0		&&	3.17$\times 10^{-4}$		&	0.0	&&	2.28$\times 10^{-4}$ 	&	0.0\\
Mn	&&	2.58$\times 10^{-4}$		&	0.0		&&	1.23$\times 10^{-4}$		&	-0.2	&&	5.59$\times 10^{-5}$		&	-0.4\\
Fe	&&	3.04$\times 10^{-2}$		&	0.0		&&	2.29$\times 10^{-2}$		&	0.0	&&	1.65$\times 10^{-2}$		&	0.0\\
Ni	&&	1.59$\times 10^{-3}$		&	0.0		&&	1.20$\times 10^{-3}$ 	&	0.0	&&	8.67$\times 10^{-4}$ 	&	0.0\\
\hline

\end{tabular}
\caption{Heavy element mixtures used in generating type 1 opacity tables. The abundances are based on \citet{2009ARA&A..47..481A} and their determination of the solar photospheric composition. }
\end{table*}

\label{lastpage}

\begin{thebibliography}{}

\expandafter\ifx\csname natexlab\endcsname\relax\def\natexlab#1{#1}\fi

\bibitem[{{Adelberger} {et~al}\mbox{.}(2011){Adelberger}, {Garc{\'{\i}}a},
  {Robertson}, {Snover}, {Balantekin}, {Heeger}, {Ramsey-Musolf}, {Bemmerer},
  {Junghans}, {Bertulani}, {Chen}, {Costantini}, {Prati}, {Couder},
  {Uberseder}, {Wiescher}, {Cyburt}, {Davids}, {Freedman}, {Gai}, {Gazit},
  {Gialanella}, {Imbriani}, {Greife}, {Hass}, {Haxton}, {Itahashi}, {Kubodera},
  {Langanke}, {Leitner}, {Leitner}, {Vetter}, {Winslow}, {Marcucci},
  {Motobayashi}, {Mukhamedzhanov}, {Tribble}, {Nollett}, {Nunes}, {Park},
  {Parker}, {Schiavilla}, {Simpson}, {Spitaleri}, {Strieder}, {Trautvetter},
  {Suemmerer}, \& {Typel}}]{2011RvMP...83..195A}
{Adelberger} E.~G. {et~al.}, 2011, Reviews of Modern Physics, 83, 195

\bibitem[{{Alcaino} {et~al}\mbox{.}(1987){Alcaino}, {Buonanno}, {Caloi},
  {Castellani}, {Corsi}, {Iannicola}, \& {Liller}}]{1987AJ.....94..917A}
{Alcaino} G., {Buonanno} R., {Caloi} V., {Castellani} V., {Corsi} C.~E.,
  {Iannicola} G., {Liller} W., 1987, \aj, 94, 917

\bibitem[{{Alongi} {et~al}\mbox{.}(1991){Alongi}, {Bertelli}, {Bressan}, \&
  {Chiosi}}]{1991A&A...244...95A}
{Alongi} M., {Bertelli} G., {Bressan} A., {Chiosi} C., 1991, \aap, 244, 95

\bibitem[{{Alonso}, {Arribas} \& {Mart{\'{\i}}nez-Roger}(1999){Alonso},
  {Arribas}, \& {Mart{\'{\i}}nez-Roger}}]{1999A&AS..140..261A}
{Alonso} A., {Arribas} S., {Mart{\'{\i}}nez-Roger} C., 1999, \aaps, 140, 261

\bibitem[{{Angelou} {et~al}\mbox{.}(2011){Angelou}, {Church}, {Stancliffe},
  {Lattanzio}, \& {Smith}}]{2011ApJ...728...79A}
{Angelou} G.~C., {Church} R.~P., {Stancliffe} R.~J., {Lattanzio} J.~C., {Smith}
  G.~H., 2011, \apj, 728, 79

\bibitem[{{Angelou} {et~al}\mbox{.}(2012){Angelou}, {Stancliffe}, {Church},
  {Lattanzio}, \& {Smith}}]{2012ApJ...749..128A}
{Angelou} G.~C., {Stancliffe} R.~J., {Church} R.~P., {Lattanzio} J.~C., {Smith}
  G.~H., 2012, \apj, 749, 128

\bibitem[{{Angulo} {et~al}\mbox{.}(1999){Angulo}, {Arnould}, {Rayet},
  {Descouvemont}, {Baye}, {Leclercq-Willain}, {Coc}, {Barhoumi}, {Aguer},
  {Rolfs}, {Kunz}, {Hammer}, {Mayer}, {Paradellis}, {Kossionides}, {Chronidou},
  {Spyrou}, {degl'Innocenti}, {Fiorentini}, {Ricci}, {Zavatarelli},
  {Providencia}, {Wolters}, {Soares}, {Grama}, {Rahighi}, {Shotter}, \& {Lamehi
  Rachti}}]{1999NuPhA.656....3A}
{Angulo} C. {et~al.}, 1999, Nuclear Physics A, 656, 3

\bibitem[{{Asplund} {et~al}\mbox{.}(2009){Asplund}, {Grevesse}, {Sauval}, \&
  {Scott}}]{2009ARA&A..47..481A}
{Asplund} M., {Grevesse} N., {Sauval} A.~J., {Scott} P., 2009, \araa, 47, 481

\bibitem[{{B{\ae}rentzen}(1965)}]{1965ZA.....62..221B}
{B{\ae}rentzen} J., 1965, \zap, 62, 221

\bibitem[{Bailey {et~al}\mbox{.}(2015)Bailey, Nagayama, Loisel, Rochau,
  Blancard, Colgan, Cosse, Faussurier, Fontes, Gilleron, Golovkin, Hansen,
  Iglesias, Kilcrease, MacFarlane, Mancini, Nahar, Orban, Pain, Pradhan,
  Sherrill, \& Wilson}]{ironopac}
Bailey J.~E. {et~al.}, 2015, Nature, 517, 56

\bibitem[{{Beaudet} \& {Tassoul}(1971)}]{1971A&A....13..209B}
{Beaudet} G., {Tassoul} M., 1971, \aap, 13, 209

\bibitem[{{Bellman} {et~al}\mbox{.}(2001){Bellman}, {Briley}, {Smith}, \&
  {Claver}}]{2001PASP..113..326B}
{Bellman} S., {Briley} M.~M., {Smith} G.~H., {Claver} C.~F., 2001, \pasp, 113,
  326

\bibitem[{{Bjork} \& {Chaboyer}(2006)}]{2006ApJ...641.1102B}
{Bjork} S.~R., {Chaboyer} B., 2006, \apj, 641, 1102

\bibitem[{{B{\"o}hm-Vitense}(1958)}]{1958ZA.....46..108B}
{B{\"o}hm-Vitense} E., 1958, \zap, 46, 108

\bibitem[{{Bressan} {et~al}\mbox{.}(2012){Bressan}, {Marigo}, {Girardi},
  {Salasnich}, {Dal Cero}, {Rubele}, \& {Nanni}}]{2012MNRAS.427..127B}
{Bressan} A., {Marigo} P., {Girardi} L., {Salasnich} B., {Dal Cero} C.,
  {Rubele} S., {Nanni} A., 2012, \mnras, 427, 127

\bibitem[{{Briley} {et~al}\mbox{.}(1990){Briley}, {Bell}, {Hoban}, \&
  {Dickens}}]{1990ApJ...359..307B}
{Briley} M.~M., {Bell} R.~A., {Hoban} S., {Dickens} R.~J., 1990, \apj, 359, 307

\bibitem[{{Brown}, {Garaud} \& {Stellmach}(2013){Brown}, {Garaud}, \&
  {Stellmach}}]{2013ApJ...768...34B}
{Brown} J.~M., {Garaud} P., {Stellmach} S., 2013, \apj, 768, 34

\bibitem[{{Campbell}({2007})}]{simthesis}
{Campbell} S., {2007}, PhD thesis, {Monash University}

\bibitem[{{Campbell} \& {Lattanzio}(2008)}]{2008A&A...490..769C}
{Campbell} S.~W., {Lattanzio} J.~C., 2008, \aap, 490, 769

\bibitem[{{Cannon}(1993)}]{1993MNRAS.263..817C}
{Cannon} R.~C., 1993, \mnras, 263, 817

\bibitem[{{Cantiello} \& {Langer}(2010)}]{2010A&A...521A...9C}
{Cantiello} M., {Langer} N., 2010, \aap, 521, A9+

\bibitem[{{Carbon} {et~al}\mbox{.}(1982){Carbon}, {Romanishin}, {Langer},
  {Butler}, {Kemper}, {Trefzger}, {Kraft}, \& {Suntzeff}}]{1982ApJS...49..207C}
{Carbon} D.~F., {Romanishin} W., {Langer} G.~E., {Butler} D., {Kemper} E.,
  {Trefzger} C.~F., {Kraft} R.~P., {Suntzeff} N.~B., 1982, \apjs, 49, 207

\bibitem[{{Carretta} {et~al}\mbox{.}(2010{\natexlab{a}}){Carretta},
  {Bragaglia}, {Gratton}, {Lucatello}, {Bellazzini}, {Catanzaro}, {Leone},
  {Momany}, {Piotto}, \& {D'Orazi}}]{2010A&A...520A..95C}
{Carretta} E. {et~al.}, 2010{\natexlab{a}}, \aap, 520, A95

\bibitem[{{Carretta} {et~al}\mbox{.}(2009){Carretta}, {Bragaglia}, {Gratton},
  {Lucatello}, {Catanzaro}, {Leone}, {Bellazzini}, {Claudi}, {D'Orazi},
  {Momany}, {Ortolani}, {Pancino}, {Piotto}, {Recio-Blanco}, \&
  {Sabbi}}]{2009A&A...505..117C}
{Carretta} E. {et~al.}, 2009, \aap, 505, 117

\bibitem[{{Carretta} {et~al}\mbox{.}(2010{\natexlab{b}}){Carretta},
  {Bragaglia}, {Gratton}, {Recio-Blanco}, {Lucatello}, {D'Orazi}, \&
  {Cassisi}}]{2010A&A...516A..55C}
{Carretta} E., {Bragaglia} A., {Gratton} R.~G., {Recio-Blanco} A., {Lucatello}
  S., {D'Orazi} V., {Cassisi} S., 2010{\natexlab{b}}, \aap, 516, A55

\bibitem[{{Cassisi}, {degl'Innocenti} \& {Salaris}(1997){Cassisi},
  {degl'Innocenti}, \& {Salaris}}]{1997MNRAS.290..515C}
{Cassisi} S., {degl'Innocenti} S., {Salaris} M., 1997, \mnras, 290, 515

\bibitem[{{Cassisi} {et~al}\mbox{.}(2011){Cassisi}, {Mar{\'{\i}}n-Franch},
  {Salaris}, {Aparicio}, {Monelli}, \& {Pietrinferni}}]{2011A&A...527A..59C}
{Cassisi} S., {Mar{\'{\i}}n-Franch} A., {Salaris} M., {Aparicio} A., {Monelli}
  M., {Pietrinferni} A., 2011, \aap, 527, A59

\bibitem[{{Cassisi} \& {Salaris}(1997)}]{1997MNRAS.285..593C}
{Cassisi} S., {Salaris} M., 1997, \mnras, 285, 593

\bibitem[{{Cassisi}, {Salaris} \& {Pietrinferni}(2013){Cassisi}, {Salaris}, \&
  {Pietrinferni}}]{2013MmSAI..84...91C}
{Cassisi} S., {Salaris} M., {Pietrinferni} A., 2013, \memsai, 84, 91

\bibitem[{{Castelli}, {Gratton} \& {Kurucz}(1997){Castelli}, {Gratton}, \&
  {Kurucz}}]{1997A&A...318..841C}
{Castelli} F., {Gratton} R.~G., {Kurucz} R.~L., 1997, \aap, 318, 841

\bibitem[{{Caughlan} \& {Fowler}(1988)}]{1988ADNDT..40..283C}
{Caughlan} G.~R., {Fowler} W.~A., 1988, Atomic Data and Nuclear Data Tables,
  40, 283

\bibitem[{{Chaboyer} {et~al}\mbox{.}(2001){Chaboyer}, {Fenton}, {Nelan},
  {Patnaude}, \& {Simon}}]{2001ApJ...562..521C}
{Chaboyer} B., {Fenton} W.~H., {Nelan} J.~E., {Patnaude} D.~J., {Simon} F.~E.,
  2001, \apj, 562, 521

\bibitem[{{Chanam{\'e}}, {Pinsonneault} \& {Terndrup}(2005){Chanam{\'e}},
  {Pinsonneault}, \& {Terndrup}}]{Chaname2005}
{Chanam{\'e}} J., {Pinsonneault} M., {Terndrup} D.~M., 2005, \apj, 631, 540

\bibitem[{{Charbonnel}(1994)}]{1994A&A...282..811C}
{Charbonnel} C., 1994, \aap, 282, 811

\bibitem[{{Charbonnel}(1995)}]{1995ApJ...453L..41C}
{Charbonnel} C., 1995, \apjl, 453, L41+

\bibitem[{{Charbonnel}, {Brown} \& {Wallerstein}(1998){Charbonnel}, {Brown}, \&
  {Wallerstein}}]{1998A&A...332..204Cgr}
{Charbonnel} C., {Brown} J.~A., {Wallerstein} G., 1998, \aap, 332, 204

\bibitem[{{Charbonnel} \& {Lagarde}(2010)}]{2010A&A...522A..10C}
{Charbonnel} C., {Lagarde} N., 2010, \aap, 522, A10+

\bibitem[{{Charbonnel} \& {Zahn}(2007{\natexlab{a}})}]{2007A&A...476L..29C}
{Charbonnel} C., {Zahn} J., 2007{\natexlab{a}}, \aap, 476, L29

\bibitem[{{Charbonnel} \& {Zahn}(2007{\natexlab{b}})}]{2007A&A...467L..15C}
{Charbonnel} C., {Zahn} J., 2007{\natexlab{b}}, \aap, 467, L15

\bibitem[{{Church} {et~al}\mbox{.}(2009){Church}, {Cristallo}, {Lattanzio},
  {Stancliffe}, {Straniero}, \& {Cannon}}]{2009PASA...26..217C}
{Church} R.~P., {Cristallo} S., {Lattanzio} J.~C., {Stancliffe} R.~J.,
  {Straniero} O., {Cannon} R.~C., 2009, \pasa, 26, 217

\bibitem[{Church {et~al}\mbox{.}(2014)Church, Lattanzio, Angelou, Tout, \&
  Stancliffe}]{Church11092014}
Church R.~P., Lattanzio J., Angelou G., Tout C.~A., Stancliffe R.~J., 2014,
  Monthly Notices of the Royal Astronomical Society, 443, 977

\bibitem[{{Constantino} {et~al}\mbox{.}(2014){Constantino}, {Campbell},
  {Gil-Pons}, \& {Lattanzio}}]{2014ApJ...784...56C}
{Constantino} T., {Campbell} S., {Gil-Pons} P., {Lattanzio} J., 2014, \apj,
  784, 56

\bibitem[{{D'Antona} \& {Caloi}(2008)}]{2008MNRAS.390..693D}
{D'Antona} F., {Caloi} V., 2008, \mnras, 390, 693

\bibitem[{{De Angeli} {et~al}\mbox{.}(2005){De Angeli}, {Piotto}, {Cassisi},
  {Busso}, {Recio-Blanco}, {Salaris}, {Aparicio}, \&
  {Rosenberg}}]{2005AJ....130..116D}
{De Angeli} F., {Piotto} G., {Cassisi} S., {Busso} G., {Recio-Blanco} A.,
  {Salaris} M., {Aparicio} A., {Rosenberg} A., 2005, \aj, 130, 116

\bibitem[{{Dearborn}, {Bolton} \& {Eggleton}(1975){Dearborn}, {Bolton}, \&
  {Eggleton}}]{1975MNRAS.170P...7D}
{Dearborn} D.~S., {Bolton} A.~J.~C., {Eggleton} P.~P., 1975, \mnras, 170, 7P

\bibitem[{{Demarque} {et~al}\mbox{.}(2000){Demarque}, {Zinn}, {Lee}, \&
  {Yi}}]{2000AJ....119.1398D}
{Demarque} P., {Zinn} R., {Lee} Y.-W., {Yi} S., 2000, \aj, 119, 1398

\bibitem[{{Denissenkov}(2010)}]{2010ApJ...723..563D}
{Denissenkov} P.~A., 2010, \apj, 723, 563

\bibitem[{{Denissenkov} \& {Merryfield}(2011)}]{2011ApJ...727L...8D}
{Denissenkov} P.~A., {Merryfield} W.~J., 2011, \apjl, 727, L8+

\bibitem[{{Denissenkov} \& {Pinsonneault}(2008)}]{2008ApJ...684..626D}
{Denissenkov} P.~A., {Pinsonneault} M., 2008, \apj, 684, 626

\bibitem[{{Denissenkov} \& {VandenBerg}(2003)}]{2003ApJ...593..509D}
{Denissenkov} P.~A., {VandenBerg} D.~A., 2003, \apj, 593, 509

\bibitem[{{Descouvemont} {et~al}\mbox{.}(2004){Descouvemont}, {Adahchour},
  {Angulo}, {Coc}, \& {Vangioni-Flam}}]{2004ADNDT..88..203D}
{Descouvemont} P., {Adahchour} A., {Angulo} C., {Coc} A., {Vangioni-Flam} E.,
  2004, Atomic Data and Nuclear Data Tables, 88, 203

\bibitem[{{Di Cecco} {et~al}\mbox{.}(2010){Di Cecco}, {Bono}, {Stetson},
  {Pietrinferni}, {Becucci}, {Cassisi}, {Degl'Innocenti}, {Iannicola}, {Prada
  Moroni}, {Buonanno}, {Calamida}, {Caputo}, {Castellani}, {Corsi}, {Ferraro},
  {Dall'Ora}, {Monelli}, {Nonino}, {Piersimoni}, {Pulone}, {Romaniello},
  {Salaris}, {Walker}, \& {Zoccali}}]{2010ApJ...712..527D}
{Di Cecco} A. {et~al.}, 2010, \apj, 712, 527

\bibitem[{{D'Orazi} {et~al}\mbox{.}(2014){D'Orazi}, {Angelou}, {Gratton},
  {Lattanzio}, {Bragaglia}, {Carretta}, {Lucatello}, \&
  {Momany}}]{2014ApJ...791...39D}
{D'Orazi} V., {Angelou} G.~C., {Gratton} R.~G., {Lattanzio} J.~C., {Bragaglia}
  A., {Carretta} E., {Lucatello} S., {Momany} Y., 2014, \apj, 791, 39

\bibitem[{{D'Orazi} {et~al}\mbox{.}(2015{\natexlab{a}}){D'Orazi}, {Gratton},
  {Angelou}, {Bragaglia}, {Carretta}, {Lattanzio}, {Lucatello}, {Momany}, \&
  {Sollima}}]{2015ApJ...801L..32D}
{D'Orazi} V. {et~al.}, 2015{\natexlab{a}}, \apjl, 801, L32

\bibitem[{{D'Orazi} {et~al}\mbox{.}(2015{\natexlab{b}}){D'Orazi}, {Gratton},
  {Angelou}, {Bragaglia}, {Carretta}, {Lattanzio}, {Lucatello}, {Momany},
  {Sollima}, \& {Beccari}}]{2015arXiv150305925D}
{D'Orazi} V. {et~al.}, 2015{\natexlab{b}}, ArXiv e-prints

\bibitem[{{Freytag}, {Ludwig} \& {Steffen}(1996){Freytag}, {Ludwig}, \&
  {Steffen}}]{1996A&A...313..497F}
{Freytag} B., {Ludwig} H.-G., {Steffen} M., 1996, \aap, 313, 497

\bibitem[{{Fusi Pecci} {et~al}\mbox{.}(1990){Fusi Pecci}, {Ferraro}, {Crocker},
  {Rood}, \& {Buonanno}}]{1990A&A...238...95F}
{Fusi Pecci} F., {Ferraro} F.~R., {Crocker} D.~A., {Rood} R.~T., {Buonanno} R.,
  1990, \aap, 238, 95

\bibitem[{{Gilroy} \& {Brown}(1991)}]{1991ApJ...371..578G}
{Gilroy} K.~K., {Brown} J.~A., 1991, \apj, 371, 578

\bibitem[{{Gonz{\'a}lez Hern{\'a}ndez} {et~al}\mbox{.}(2009){Gonz{\'a}lez
  Hern{\'a}ndez}, {Bonifacio}, {Caffau}, {Steffen}, {Ludwig}, {Behara},
  {Sbordone}, {Cayrel}, \& {Zaggia}}]{2009AA...505L..13G}
{Gonz{\'a}lez Hern{\'a}ndez} J.~I. {et~al.}, 2009, \aap, 505, L13

\bibitem[{{Gratton} {et~al}\mbox{.}(2001){Gratton}, {Bonifacio}, {Bragaglia},
  {Carretta}, {Castellani}, {Centurion}, {Chieffi}, {Claudi}, {Clementini},
  {D'Antona}, {Desidera}, {Fran{\c c}ois}, {Grundahl}, {Lucatello}, {Molaro},
  {Pasquini}, {Sneden}, {Spite}, \& {Straniero}}]{2001A&A...369...87G}
{Gratton} R.~G. {et~al.}, 2001, \aap, 369, 87

\bibitem[{{Gratton}, {Carretta} \& {Bragaglia}(2012){Gratton}, {Carretta}, \&
  {Bragaglia}}]{2012A&ARv..20...50G}
{Gratton} R.~G., {Carretta} E., {Bragaglia} A., 2012, \aapr, 20, 50

\bibitem[{{Gratton} {et~al}\mbox{.}(2010){Gratton}, {Carretta}, {Bragaglia},
  {Lucatello}, \& {D'Orazi}}]{2010A&A...517A..81G}
{Gratton} R.~G., {Carretta} E., {Bragaglia} A., {Lucatello} S., {D'Orazi} V.,
  2010, \aap, 517, A81

\bibitem[{{Gratton} {et~al}\mbox{.}(2012){Gratton}, {Lucatello}, {Carretta},
  {Bragaglia}, {D'Orazi}, {Al Momany}, {Sollima}, {Salaris}, \&
  {Cassisi}}]{2012A&A...539A..19G}
{Gratton} R.~G. {et~al.}, 2012, \aap, 539, A19

\bibitem[{{Gratton} {et~al}\mbox{.}(2011){Gratton}, {Lucatello}, {Carretta},
  {Bragaglia}, {D'Orazi}, \& {Momany}}]{2011A&A...534A.123G}
{Gratton} R.~G., {Lucatello} S., {Carretta} E., {Bragaglia} A., {D'Orazi} V.,
  {Momany} Y.~A., 2011, \aap, 534, A123

\bibitem[{{Gratton} {et~al}\mbox{.}(2000){Gratton}, {Sneden}, {Carretta}, \&
  {Bragaglia}}]{2000A&A...354..169G}
{Gratton} R.~G., {Sneden} C., {Carretta} E., {Bragaglia} A., 2000, \aap, 354,
  169

\bibitem[{{Grevesse} \& {Noels}(1993)}]{1993oee..conf...15G}
{Grevesse} N., {Noels} A., 1993, in Origin and Evolution of the Elements,
  {Prantzos} N., {Vangioni-Flam} E., {Casse} M., eds., pp. 15--25

\bibitem[{{Gruyters} {et~al}\mbox{.}(2013){Gruyters}, {Korn}, {Richard},
  {Grundahl}, {Collet}, {Mashonkina}, {Osorio}, \&
  {Barklem}}]{2013A&A...555A..31G}
{Gruyters} P., {Korn} A.~J., {Richard} O., {Grundahl} F., {Collet} R.,
  {Mashonkina} L.~I., {Osorio} Y., {Barklem} P.~S., 2013, \aap, 555, A31

\bibitem[{{Harris} {et~al}\mbox{.}(1983){Harris}, {Fowler}, {Caughlan}, \&
  {Zimmerman}}]{1983ARAA..21..165H}
{Harris} M.~J., {Fowler} W.~A., {Caughlan} G.~R., {Zimmerman} B.~A., 1983,
  \araa, 21, 165

\bibitem[{{Harris}(1996)}]{1996AJ....112.1487H}
{Harris} W.~E., 1996, \aj, 112, 1487

\bibitem[{{Herwig} {et~al}\mbox{.}(1997){Herwig}, {Bloecker}, {Schoenberner},
  \& {El Eid}}]{1997A&A...324L..81H}
{Herwig} F., {Bloecker} T., {Schoenberner} D., {El Eid} M., 1997, \aap, 324,
  L81

\bibitem[{{Houdashelt}, {Bell} \& {Sweigart}(2000){Houdashelt}, {Bell}, \&
  {Sweigart}}]{2000AJ....119.1448H}
{Houdashelt} M.~L., {Bell} R.~A., {Sweigart} A.~V., 2000, \aj, 119, 1448

\bibitem[{{Iben}(1968)}]{1968Natur.220..143I}
{Iben} I., 1968, \nat, 220, 143

\bibitem[{{Iben}(1967)}]{1967ApJ...147..624I}
{Iben}, Jr. I., 1967, \apj, 147, 624

\bibitem[{{Iben} \& {Faulkner}(1968)}]{1968ApJ...153..101I}
{Iben}, Jr. I., {Faulkner} J., 1968, \apj, 153, 101

\bibitem[{{Iglesias} \& {Rogers}(1996)}]{1996ApJ...464..943I}
{Iglesias} C.~A., {Rogers} F.~J., 1996, \apj, 464, 943

\bibitem[{{Johnson} \& {Pilachowski}(2010)}]{2010ApJ...722.1373J}
{Johnson} C.~I., {Pilachowski} C.~A., 2010, \apj, 722, 1373

\bibitem[{{King}, {Da Costa} \& {Demarque}(1985){King}, {Da Costa}, \&
  {Demarque}}]{1985ApJ...299..674K}
{King} C.~R., {Da Costa} G.~S., {Demarque} P., 1985, \apj, 299, 674

\bibitem[{{Kippenhahn}, {Ruschenplatt} \& {Thomas}(1980){Kippenhahn},
  {Ruschenplatt}, \& {Thomas}}]{1980A&A....91..175K}
{Kippenhahn} R., {Ruschenplatt} G., {Thomas} H., 1980, \aap, 91, 175

\bibitem[{{Langer} {et~al}\mbox{.}(1986){Langer}, {Kraft}, {Carbon}, {Friel},
  \& {Oke}}]{1986PASP...98..473L}
{Langer} G.~E., {Kraft} R.~P., {Carbon} D.~F., {Friel} E., {Oke} J.~B., 1986,
  \pasp, 98, 473

\bibitem[{{Lattanzio} {et~al}\mbox{.}(2015){Lattanzio}, {Siess}, {Church},
  {Angelou}, {Stancliffe}, {Doherty}, {Stephen}, \&
  {Campbell}}]{2015MNRAS.446.2673L}
{Lattanzio} J.~C., {Siess} L., {Church} R.~P., {Angelou} G., {Stancliffe}
  R.~J., {Doherty} C.~L., {Stephen} T., {Campbell} S.~W., 2015, \mnras, 446,
  2673

\bibitem[{{Lederer} \& {Aringer}(2009)}]{2009A&A...494..403L}
{Lederer} M.~T., {Aringer} B., 2009, \aap, 494, 403

\bibitem[{{Lind} {et~al}\mbox{.}(2008){Lind}, {Korn}, {Barklem}, \&
  {Grundahl}}]{2008A&A...490..777L}
{Lind} K., {Korn} A.~J., {Barklem} P.~S., {Grundahl} F., 2008, \aap, 490, 777

\bibitem[{{Lind} {et~al}\mbox{.}(2009){Lind}, {Primas}, {Charbonnel},
  {Grundahl}, \& {Asplund}}]{2009AA...503..545L}
{Lind} K., {Primas} F., {Charbonnel} C., {Grundahl} F., {Asplund} M., 2009,
  \aap, 503, 545

\bibitem[{{Magic}, {Weiss} \& {Asplund}(2015){Magic}, {Weiss}, \&
  {Asplund}}]{2015A&A...573A..89M}
{Magic} Z., {Weiss} A., {Asplund} M., 2015, \aap, 573, A89

\bibitem[{{Marino} {et~al}\mbox{.}(2009){Marino}, {Milone}, {Piotto},
  {Villanova}, {Bedin}, {Bellini}, \& {Renzini}}]{2009A&A...505.1099M}
{Marino} A.~F., {Milone} A.~P., {Piotto} G., {Villanova} S., {Bedin} L.~R.,
  {Bellini} A., {Renzini} A., 2009, \aap, 505, 1099

\bibitem[{{Marino} {et~al}\mbox{.}(2014){Marino}, {Milone}, {Przybilla},
  {Bergemann}, {Lind}, {Asplund}, {Cassisi}, {Catelan}, {Casagrande},
  {Valcarce}, {Bedin}, {Cort{\'e}s}, {D'Antona}, {Jerjen}, {Piotto},
  {Schlesinger}, {Zoccali}, \& {Angeloni}}]{2014MNRAS.437.1609M}
{Marino} A.~F. {et~al.}, 2014, \mnras, 437, 1609

\bibitem[{{Martell}, {Smith} \& {Briley}(2008){Martell}, {Smith}, \&
  {Briley}}]{2008AJ....136.2522M}
{Martell} S.~L., {Smith} G.~H., {Briley} M.~M., 2008, \aj, 136, 2522

\bibitem[{{Meissner} \& {Weiss}(2006)}]{2006A&A...456.1085M}
{Meissner} F., {Weiss} A., 2006, \aap, 456, 1085

\bibitem[{{Michaud}, {Richer} \& {Richard}(2010){Michaud}, {Richer}, \&
  {Richard}}]{2010A&A...510A.104M}
{Michaud} G., {Richer} J., {Richard} O., 2010, \aap, 510, A104

\bibitem[{{Milone} {et~al}\mbox{.}(2012){Milone}, {Marino}, {Cassisi},
  {Piotto}, {Bedin}, {Anderson}, {Allard}, {Aparicio}, {Bellini}, {Buonanno},
  {Monelli}, \& {Pietrinferni}}]{2012ApJ...754L..34M}
{Milone} A.~P. {et~al.}, 2012, \apjl, 754, L34

\bibitem[{{Milone} {et~al}\mbox{.}(2014){Milone}, {Marino}, {Dotter}, {Norris},
  {Jerjen}, {Piotto}, {Cassisi}, {Bedin}, {Recio Blanco}, {Sarajedini},
  {Asplund}, {Monelli}, \& {Aparicio}}]{2014ApJ...785...21M}
{Milone} A.~P. {et~al.}, 2014, \apj, 785, 21

\bibitem[{{Mucciarelli} {et~al}\mbox{.}(2011){Mucciarelli}, {Salaris},
  {Lovisi}, {Ferraro}, {Lanzoni}, {Lucatello}, \&
  {Gratton}}]{2011MNRAS.412...81M}
{Mucciarelli} A., {Salaris} M., {Lovisi} L., {Ferraro} F.~R., {Lanzoni} B.,
  {Lucatello} S., {Gratton} R.~G., 2011, \mnras, 412, 81

\bibitem[{{Nataf} {et~al}\mbox{.}(2013){Nataf}, {Gould}, {Pinsonneault}, \&
  {Udalski}}]{2013ApJ...766...77N}
{Nataf} D.~M., {Gould} A.~P., {Pinsonneault} M.~H., {Udalski} A., 2013, \apj,
  766, 77

\bibitem[{{Palacios} {et~al}\mbox{.}(2006){Palacios}, {Charbonnel}, {Talon}, \&
  {Siess}}]{2006A&A...453..261P}
{Palacios} A., {Charbonnel} C., {Talon} S., {Siess} L., 2006, \aap, 453, 261

\bibitem[{{Palmerini} {et~al}\mbox{.}(2011){Palmerini}, {La Cognata},
  {Cristallo}, \& {Busso}}]{2011ApJ...729....3P}
{Palmerini} S., {La Cognata} M., {Cristallo} S., {Busso} M., 2011, \apj, 729, 3

\bibitem[{{Pietrinferni} {et~al}\mbox{.}(2004){Pietrinferni}, {Cassisi},
  {Salaris}, \& {Castelli}}]{2004ApJ...612..168P}
{Pietrinferni} A., {Cassisi} S., {Salaris} M., {Castelli} F., 2004, \apj, 612,
  168

\bibitem[{{Piotto} {et~al}\mbox{.}(2007){Piotto}, {Bedin}, {Anderson}, {King},
  {Cassisi}, {Milone}, {Villanova}, {Pietrinferni}, \&
  {Renzini}}]{2007ApJ...661L..53P}
{Piotto} G. {et~al.}, 2007, \apjl, 661, L53

\bibitem[{{Placco} {et~al}\mbox{.}(2014){Placco}, {Frebel}, {Beers}, \&
  {Stancliffe}}]{2014ApJ...797...21P}
{Placco} V.~M., {Frebel} A., {Beers} T.~C., {Stancliffe} R.~J., 2014, \apj,
  797, 21

\bibitem[{{Planck Collaboration} {et~al}\mbox{.}(2014){Planck Collaboration},
  {Ade}, {Aghanim}, {Alves}, {Armitage-Caplan}, {Arnaud}, {Ashdown},
  {Atrio-Barandela}, {Aumont}, {Aussel}, \& et~al.}]{2014A&A...571A...1P}
{Planck Collaboration} {et~al.}, 2014, \aap, 571, A1

\bibitem[{{Richard}, {Michaud} \& {Richer}(2005){Richard}, {Michaud}, \&
  {Richer}}]{2005ApJ...619..538R}
{Richard} O., {Michaud} G., {Richer} J., 2005, \apj, 619, 538

\bibitem[{{Riello} {et~al}\mbox{.}(2003){Riello}, {Cassisi}, {Piotto},
  {Recio-Blanco}, {De Angeli}, {Salaris}, {Pietrinferni}, {Bono}, \&
  {Zoccali}}]{2003A&A...410..553R}
{Riello} M. {et~al.}, 2003, \aap, 410, 553

\bibitem[{{Rogers} \& {Nayfonov}(2002)}]{2002ApJ...576.1064R}
{Rogers} F.~J., {Nayfonov} A., 2002, \apj, 576, 1064

\bibitem[{{Salaris} \& {Cassisi}(2015)}]{2015arXiv150304582S}
{Salaris} M., {Cassisi} S., 2015, ArXiv e-prints

\bibitem[{{Salaris}, {Cassisi} \& {Weiss}(2002){Salaris}, {Cassisi}, \&
  {Weiss}}]{2002PASP..114..375S}
{Salaris} M., {Cassisi} S., {Weiss} A., 2002, \pasp, 114, 375

\bibitem[{{Salaris}, {Chieffi} \& {Straniero}(1993){Salaris}, {Chieffi}, \&
  {Straniero}}]{1993ApJ...414..580S}
{Salaris} M., {Chieffi} A., {Straniero} O., 1993, \apj, 414, 580

\bibitem[{{Salaris} {et~al}\mbox{.}(2006){Salaris}, {Weiss}, {Ferguson}, \&
  {Fusilier}}]{2006ApJ...645.1131S}
{Salaris} M., {Weiss} A., {Ferguson} J.~W., {Fusilier} D.~J., 2006, \apj, 645,
  1131

\bibitem[{{Shetrone} {et~al}\mbox{.}(2010){Shetrone}, {Martell}, {Wilkerson},
  {Adams}, {Siegel}, {Smith}, \& {Bond}}]{2010AJ....140.1119S}
{Shetrone} M., {Martell} S.~L., {Wilkerson} R., {Adams} J., {Siegel} M.~H.,
  {Smith} G.~H., {Bond} H.~E., 2010, \aj, 140, 1119

\bibitem[{{Shetrone}(2003)}]{2003ApJ...585L..45S}
{Shetrone} M.~D., 2003, \apjl, 585, L45

\bibitem[{{Smith} \& {Martell}(2003)}]{2003PASP..115.1211S}
{Smith} G.~H., {Martell} S.~L., 2003, \pasp, 115, 1211

\bibitem[{{Spite} \& {Spite}(1982)}]{1982A&A...115..357S}
{Spite} F., {Spite} M., 1982, \aap, 115, 357

\bibitem[{{Stancliffe}(2010)}]{2010MNRAS.403..505S}
{Stancliffe} R.~J., 2010, \mnras, 403, 505

\bibitem[{{Stancliffe} {et~al}\mbox{.}(2009){Stancliffe}, {Church}, {Angelou},
  \& {Lattanzio}}]{2009MNRAS.396.2313S}
{Stancliffe} R.~J., {Church} R.~P., {Angelou} G.~C., {Lattanzio} J.~C., 2009,
  \mnras, 396, 2313

\bibitem[{{St{\"o}kl}(2008)}]{2008A&A...490.1181S}
{St{\"o}kl} A., 2008, \aap, 490, 1181

\bibitem[{{Sweigart} \& {Mengel}(1979)}]{1979ApJ...229..624S}
{Sweigart} A.~V., {Mengel} J.~G., 1979, \apj, 229, 624

\bibitem[{{Timmes} \& {Arnett}(1999)}]{1999ApJS..125..277T}
{Timmes} F.~X., {Arnett} D., 1999, \apjs, 125, 277

\bibitem[{{Timmes} \& {Swesty}(2000)}]{2000ApJS..126..501T}
{Timmes} F.~X., {Swesty} F.~D., 2000, \apjs, 126, 501

\bibitem[{{Tomkin}, {Luck} \& {Lambert}(1976){Tomkin}, {Luck}, \&
  {Lambert}}]{1976ApJ...210..694T}
{Tomkin} J., {Luck} R.~E., {Lambert} D.~L., 1976, \apj, 210, 694

\bibitem[{{Traxler}, {Garaud} \& {Stellmach}(2011){Traxler}, {Garaud}, \&
  {Stellmach}}]{2011ApJ...728L..29T}
{Traxler} A., {Garaud} P., {Stellmach} S., 2011, \apjl, 728, L29+

\bibitem[{{Troisi} {et~al}\mbox{.}(2011){Troisi}, {Bono}, {Stetson},
  {Pietrinferni}, {Weiss}, {Fabrizio}, {Ferraro}, {Cecco}, {Iannicola},
  {Buonanno}, {Calamida}, {Caputo}, {Corsi}, {Dall'Ora}, {Kunder}, {Monelli},
  {Nonino}, {Piersimoni}, {Pulone}, {Romaniello}, {Walker}, \&
  {Zoccali}}]{2011PASP..123..879T}
{Troisi} F. {et~al.}, 2011, \pasp, 123, 879

\bibitem[{{Ulrich}(1972)}]{1972ApJ...172..165U}
{Ulrich} R.~K., 1972, \apj, 172, 165

\bibitem[{{VandenBerg}, {Bergbusch} \& {Dowler}(2006){VandenBerg}, {Bergbusch},
  \& {Dowler}}]{2006ApJS..162..375V}
{VandenBerg} D.~A., {Bergbusch} P.~A., {Dowler} P.~D., 2006, \apjs, 162, 375

\bibitem[{{Weiss} \& {Charbonnel}(2004)}]{2004MmSAI..75..347W}
{Weiss} A., {Charbonnel} C., 2004, Memorie della Societa Astronomica Italiana,
  75, 347

\bibitem[{{Yong} {et~al}\mbox{.}(2014){Yong}, {Roederer}, {Grundahl}, {Da
  Costa}, {Karakas}, {Norris}, {Aoki}, {Fishlock}, {Marino}, {Milone}, \&
  {Shingles}}]{2014MNRAS.441.3396Y}
{Yong} D. {et~al.}, 2014, \mnras, 441, 3396

\bibitem[{{Zoccali} {et~al}\mbox{.}(1999){Zoccali}, {Cassisi}, {Piotto},
  {Bono}, \& {Salaris}}]{1999ApJ...518L..49Z}
{Zoccali} M., {Cassisi} S., {Piotto} G., {Bono} G., {Salaris} M., 1999, \apjl,
  518, L49

\end{thebibliography}
\end{document}